\documentclass[
reprint,
showpacs,
preprintnumbers,
nobibnotes,
amsmath,amssymb,
aps,
pra,
floatfix
]{revtex4-2}
\usepackage{makecell}
\usepackage{graphicx}
\usepackage{booktabs}
\usepackage{verbatim}
\usepackage{dcolumn}
\usepackage{multirow}
\usepackage{bm}
\usepackage{nicefrac}
\usepackage{amsmath,amssymb,mathtools}
\usepackage{leftindex}
\usepackage[table]{xcolor}
\usepackage[colorlinks=true, linkcolor=blue, urlcolor=blue, citecolor=blue]{hyperref}

\newcommand{\qvec}[1]{\bm{#1}}

\begin{document}

\preprint{APS/123-QED}

\title{Discretized Halbach spheres: Icosahedral symmetry for optimal field homogeneity}
\author{Ingo Rehberg}
\email{Ingo.Rehberg@uni-bayreuth.de}
\affiliation{Institute of Physics, University of Bayreuth, 95440 Bayreuth, Germany}
\author{Peter Blümler}
\email{bluemler@uni-mainz.de}
\affiliation{Institute of Physics, University of Mainz, 55128 Mainz, Germany}
\date{\today}

\begin{abstract}
Halbach spheres provide a theoretically elegant means of generating highly homogeneous magnetic fields, but practical implementation is hindered by challenging fabrication and restricted interior access. This study examines discrete spherical Halbach configurations assembled from permanent magnets placed at the vertices of Platonic and Archimedean solids. Analytical calculations, numerical field simulations, and experimental measurements indicate that polyhedra with icosahedral symmetry achieve the most favorable balance among field strength, homogeneity, and interior accessibility.
They produce exceptionally flat fourth-order central saddle points, resulting in a usable homogeneous field volume up to a factor of 260 larger than that of traditional Halbach disk or cylindrical arrays. Several magnet assemblies composed of cubical NdFeB magnets are fabricated and their three dimensional field distributions characterized, demonstrating homogeneous regions of up to several cubic centimeters with deviations below $1\,\%$. The findings establish discrete icosahedrally symmetric magnet arrays as practical, scalable building blocks for compact, highly homogeneous magnetic field sources suited to mobile magnetic resonance, and magnetophoretic applications.
\end{abstract}

\maketitle

\section{\label{sec:intro}Introduction}
The generation of strong and highly homogeneous magnetic fields using permanent magnets is of great interest both for laboratory instrumentation and a wide range of technical applications~\cite{Coey2002,Raich2004,Coey_book,BluCasa2015,Soltner2023,Bakenecker2020,BluGuid2021}. Building on the concept of single-sided fluxes~\cite{Mallinson1973}, Klaus Halbach demonstrated a remarkably elegant solution to this challenge: an infinitely long cylindrical shell of permanent-magnet material in which the magnetization direction rotates as twice the azimuthal angle~\cite{Halbach1980}. This continuously varying magnetization produces a perfectly uniform internal field, while suppressing the external stray field.

Shortly after Halbach’s seminal contribution, several authors extended the idea from cylindrical to spherical geometries~\cite{Zijlstra1985, Leupold1987, Leupold2000}. In these spherical analogs of the Halbach cylinder, the magnetization pattern corresponds to rotating the circular cross-section of the cylinder around the field axis, thereby creating a three dimensional permanent-magnet structure capable of generating an ideal internal field. Early discussions emphasized the limited accessibility of the enclosed cavity, but subsequent theoretical work showed that equatorial and polar openings are feasible~\cite{Leupold1994} and that planes can be identified to open the ideal Halbach sphere without force~\cite{patent_spheres, BluCasa2015}. 

Despite these advances, the considerable complexity of fabrication has so far restricted practical realizations to only a few exemplary constructions, which emphasized the achievement of high field strength rather than optimal field homogeneity \cite{Bloch1998}.

The ideal Halbach cylinder cannot be realized for two fundamental reasons: (a) it is infinitely long and (b) it requires a continuously varying magnetization that cannot be manufactured. The first limitation has been addressed in practice by constructing finite-length assemblies from discrete permanent magnets, typically using planar Halbach disks that are stacked into “sandwich” configurations to better approximate the desired field~\cite{Soltner2010, Soltner2023}, or by incorporating suitable end caps to reduce field drop-off~\cite{Leupold_patent, Bjork2008}. A complementary strategy mitigates finite-length inhomogeneities through analytical corrections based on point-dipole models that depart from Halbach’s original formulation~\cite{Rehberg2025}. In contrast, a Halbach sphere overcomes the finite-length problem intrinsically, achieving an ideal internal field without compromising the underlying design principle.\\
\indent The second problem is typically addressed by discretizing the magnetic shell into segments, each with a uniform magnetization direction. This can be achieved by constructing the magnet from individually shaped and magnetized pieces, by using identically shaped magnets that are mechanically rotated in the required orientations~\cite{Raich2004}, or by magnetically aligning cylindrical magnets within supportive mounting structures~\cite{Wickenbrock2021}.

In this work, a variant of the second concept is adopted by arranging discrete permanent magnets on the surfaces of spherical shells to approximate the continuously varying magnetization of an ideal Halbach sphere. For selecting the placement and orientation of individual magnetic dipoles, the vertices of the Platonic solids provide a natural and symmetric set of candidate positions. The tetrahedral, octahedral and icosahedral symmetry classes $T_d, O_h,$ and $I_h$ each yield a distinct discretization of the spherical surface and thus a unique approximation of the ideal magnetization pattern. Among these, configurations based on icosahedral symmetry turn out to show the most favorable balance between constructability and magnetic-field uniformity, producing superior homogeneity within the spherical interior compared with other polyhedral geometries including the thirteen Archimedean solids. From this set, the “truncated icosahedron” and "truncated icosidodecahedron” were identified as particularly promising candidates for practical studies requiring homogeneous fields. Their construction is presented here.

\section{\label{sec:theory}Theory}
A complete list of symbols used in this work is provided in the Supplementary Material.

To maximize the magnetic field strength achievable with permanent magnets, it is natural to consider assemblies composed of a large number of individual elements. In order to establish upper bounds for such configurations, it is useful to consider an idealized, continuous distribution of magnetization in space.

\subsection{\label{sec:continua} Field produced by a continuous magnetization}
The field at the origin caused by a magnetization $\mathbf{M}(\mathbf{r})$ is given by
\begin{equation}
\mathbf{B}(\mathbf{0}) = \frac{\mu_0}{4\pi} \int_{\mathrm{Vol}} \left( \frac{3 (\mathbf{M}(\mathbf{r}) \cdot \hat{\mathbf{r}}) \hat{\mathbf{r}} - \mathbf{M}(\mathbf{r})}{|\mathbf{r}|^3} \right) \, d^3r,
\label{eq:continua}
\end{equation}
where the hat denotes a unity vector.
For a compact presentation of the results, it is convenient to express the magnetization as 
\begin{equation}
\mathbf{M}= \frac{B_\mathrm{R}}{\mu_0} \hat{\mathbf{M}},
\label{eq:B_r}
\end{equation}
where $B_\mathrm{R}$ denotes the remanence of the magnetized material. 

The magnetic field along a given axis is maximized when the magnetization orientation angle $\alpha_\mathrm{o}$ measured with respect to that axis, depends on the position angle $\theta_\mathrm{p}$ as ~\cite{Zijlstra1985, RehbergBlümler2025}
\begin{equation}
\tan\alpha_\mathrm{o}=\frac{3 \sin\theta_\mathrm{p}\cos\theta_\mathrm{p}}{3 \cos^2\!\theta_\mathrm{p} - 1},
\label{eq:focused}
\end{equation}
where both $\alpha_\mathrm{o}$ and $\theta_\mathrm{p}$ are defined in the plane spanned by the position vector $\qvec{r}$ and the orientation axis. Equation~(\ref{eq:focused}) defines the focused orientation ~\cite{RehbergBlümler2025}. That is set in contrast here to the classical Halbach orientation defined by
\begin{equation}
\alpha_\mathrm{o}=2\theta_\mathrm{p}.
\label{eq:Halbach}
\end{equation}
To obtain the field in the center of a \emph{thin} ring, where the height $h$ is small compared to the inner radius $R_\mathrm{in}$, the integration of Eq.~(\ref{eq:continua}) within the volume of the cylindrical disk with an outer radius $R_\mathrm{out}$ can be simplified (see Appendix~\ref{app:continua} for details). For $h \ll R_\mathrm{in}$, and for the Halbach configuration Eq.~(\ref{eq:Halbach}), this results in a magnitude of the field at the center of
\begin{equation}
^\mathrm{H}B_\mathrm{c} = B_\mathrm{R}\cdot \frac{3}{4} \cdot  
            \left [\frac{h}{R_\mathrm{in}}-\frac{h}{R_\mathrm{out}}\right].
\label{eq:HBc}
\end{equation}

\begin{figure}[ht]
\includegraphics[width=.48\textwidth]{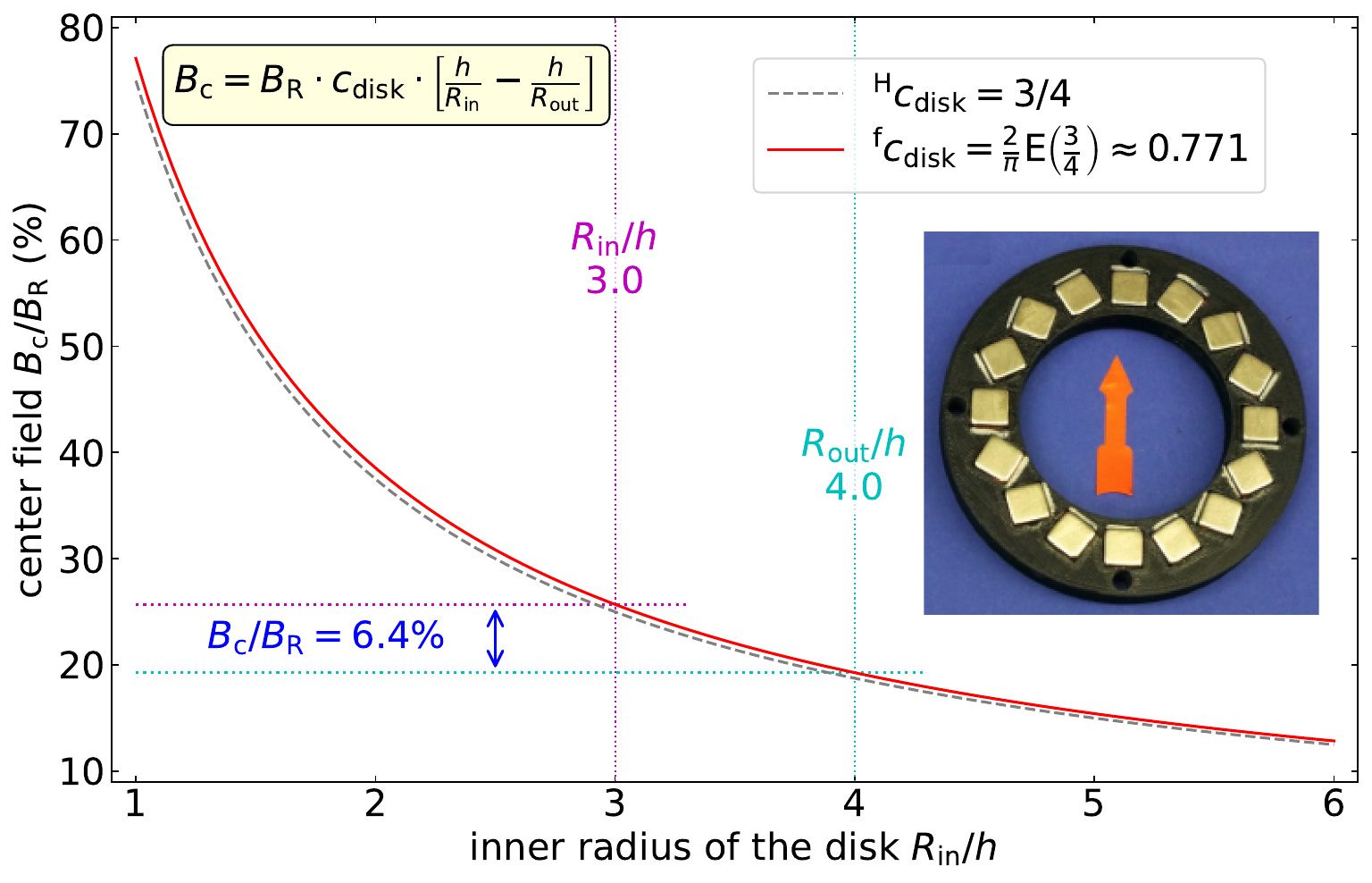}
\caption{Magnetic field at the center of a thin magnetized ring with $R_{\mathrm{out}} \to \infty$: The solid red line shows the theoretical prediction for the focused configuration. The gray dashed line corresponds to the Halbach configuration. The associated scaling factors are given in the legend. The cyan and magenta dashed lines illustrate the application of the theory to the experimental example shown in the inset.}
\label{fig:1}
\end{figure}

For the focused configuration, integration over $\phi$ and $r$ leads to a larger field
\begin{equation}
^\mathrm{f}B_\mathrm{c} = B_\mathrm{R}\cdot \frac{2}{\pi}\;\mathrm{E}\left(\frac{3}{4}\right)\cdot  
            \left [\frac{h}{R_\mathrm{in}}-\frac{h}{R_\mathrm{out}}\right],
\label{eq:fBc}
\end{equation}
where $\mathrm{E}$ is the elliptic integral of the second kind. The resulting prefactor exceeds the Halbach value of $3/4$ by $2.8\%$. Equations~(\ref{eq:HBc}) and (\ref{eq:fBc}) are shown in Fig.~\ref{fig:1} for an infinitely extended disk, i.~e. in the limit $R_{\mathrm{out}} \to \infty$. The application of the idealized formula (\ref{eq:fBc}) to an experimental realization is shown in the inset and is discussed in Appendix~\ref{app:finite ring}.

\begin{figure}[ht]
\includegraphics[width=.48\textwidth]{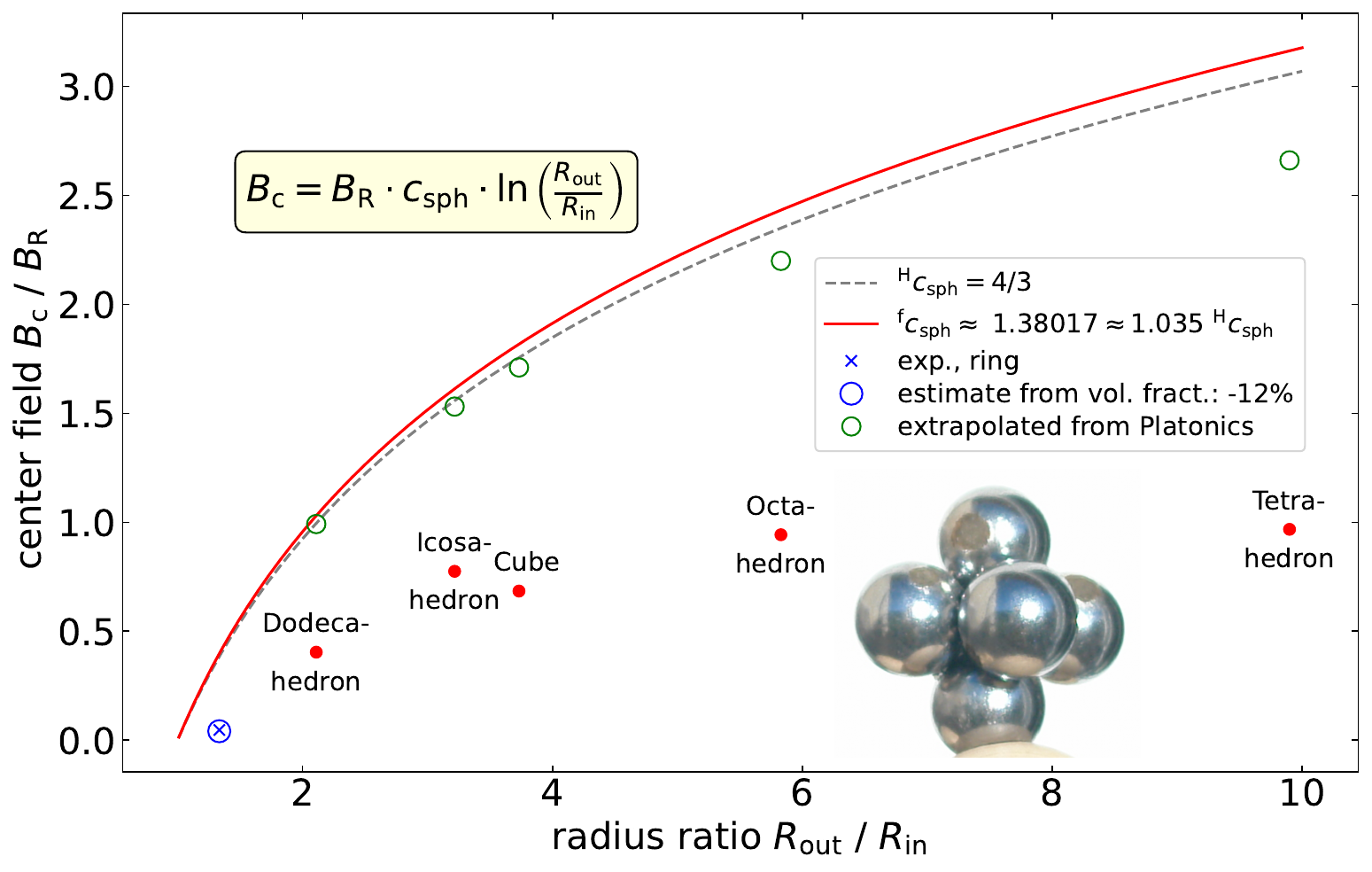}
\caption{Magnetic field at the center of a magnetized spherical shell. The solid red line shows the theoretical prediction for the focused configuration, while the gray dashed line corresponds to the Halbach configuration (see Eq.~\ref{eq:sph_th}) The associated scaling factors are given in the legend. Red dots indicate the central field $B_\mathrm{c}$ obtained for Platonic arrangements of touching magnetic spheres; an octahedral configuration is shown in the inset as an experimental example. Open green circles represent the application of the continuum theory to the five Platonic arrangements. The blue cross and open circle illustrate the application of the theory to the experimental example shown in Fig.~\ref{fig:1}.}
\label{fig:2}
\end{figure}

Figure~\ref{fig:1} clearly shows that extended disk arrangements cannot realistically produce central fields exceeding the remanence (i.~e., $B_\mathrm{c} < B_\mathrm{R}$). Achieving higher field strengths therefore requires exploiting the third dimension. In particular, distributing the magnets over spherical shells leads to a more favorable scaling: the resulting magnetic field of the three dimensional integral increases logarithmically with the outer radius of the shell~\cite{Zijlstra1985}, as illustrated in Fig.~\ref{fig:2}.

For a continuous magnetization distribution on a spherical shell, the magnetic field at the center is then given by
\begin{equation}
\label{eq:sph_th}
B_\mathrm{c} = B_\mathrm{R}\cdot c_\mathrm{sph} \cdot  
            \ln \left (\frac{R_\mathrm{out}}{R_\mathrm{in}}\right),
\end{equation}
where the prefactor for the Halbach configuration is $^\mathrm{H}c_\mathrm{sph}=4/3$.  For the focused configuration, the corresponding factor is determined numerically by integrating Eq.~(\ref{eq:continua}) over the spherical coordinates. It is found to be $^\mathrm{f}c_\mathrm{sph} \approx 1.38017$. This corresponds to an increase in the central field of approximately $3.5\%$ relative to the Halbach case. However, a key advantage of the Halbach configuration is that it yields a perfectly homogeneous magnetic field for a continuously magnetized spherical shell~\cite{Zijlstra1985}.

As discussed in Section 
\ref{sec:intro}, closed spherical surfaces with a continuous magnetization distribution are not practically realizable; moreover, they are of limited practical use because the enclosed inner volume with the strong and homogeneous field is not accessible. A feasible alternative, and the focus of this paper, is therefore to approximate such configurations by arranging discrete permanent magnets on the surface of a sphere. In this context, the vertices of Platonic solids provide highly symmetric and thus natural locations for magnetic dipoles \cite{Coxeter_book1973}. The relevant properties of these solids are summarized in Table~\ref{tab:solids}. The calculation of the center field is provided in the Appendix~\ref{sec:appC}, and additional practical details are presented in the Supplementary Material.

\begin{table}[htbp]
    \centering
    \caption{Magnetic properties of regular (Platonic) and semiregular (Archimedean) polyhedral configurations. The numbers of faces, vertices, and edges are denoted by $F$, $V$, and $E$, respectively. The polyhedra belong to three rotational symmetry groups (Sy), where $O$ and $I$ denote the chiral (rotation-only) octahedral and icosahedral symmetry groups. The leading-order term of $(B(\rho)-B_\mathrm{c})$ is listed together with the direction of the strongest field variation (ORD) and its characteristic length scale $\lambda$. The lengths ($\lambda$ and radius $R$) are given in units of $r_\mathrm{s}$, defined as half the edge length in the limit of touching magnets. The central magnetic field $B_\mathrm{c}$ produced by spherical magnets in contact is normalized to the remanence $B_\mathrm{R}$. Configurations exhibiting the connection between icosahedral symmetry and fourth-order extrema are highlighted in red, while experimentally realized configurations are indicated in yellow.} 
    \label{tab:solids}
    \begin{tabular}{@{}lcccccccc@{}}
\toprule
\textbf{Name} & $\boldsymbol{F}$ & $\boldsymbol{V}$ & $\boldsymbol{E}$ & \textbf{Sy} & \textbf{Ord}& $\boldsymbol{\lambda}/r_\text{s}$ & $\boldsymbol{R}/r_\text{s}$ & $\boldsymbol{\frac{B_\mathrm{c}}{B_\mathrm{R}}}\%$ \\
\midrule
\makecell[l]{\setlength{\tabcolsep}{0pt} Tetrahedron} & 4 & 4 & 6 & $T_d$ & $z^{1}$ & 1.9 & 1.2 & 97 \\
\makecell[l]{\setlength{\tabcolsep}{0pt} Cube \\~~ (Hexahedron)} & 6 & 8 & 12 & $O_h$ & $x^{2}$ & 2.4 & 1.7 & 68 \\
\makecell[l]{\setlength{\tabcolsep}{0pt} Octahedron} & 8 & 6 & 12 & $O_h$ & $y^{2}$ & 1.6 & 1.4 & 94 \\
\makecell[l]{\setlength{\tabcolsep}{0pt} Dodecahedron} & 12 & 20 & 30 & $\cellcolor{red!15} I_h$ & \cellcolor{red!15} $z^{4}$ & 4.2 & 2.8 & 40 \\
\rowcolor{yellow!40}\makecell[l]{\setlength{\tabcolsep}{0pt} Icosahedron} & 20 & 12 & 30 & $\cellcolor{red!15} I_h$ & \cellcolor{red!15} $y^{4}$ & 2.1 & 1.9 & 77 \\
\makecell[l]{\setlength{\tabcolsep}{0pt} Truncated \\~~ Tetrahedron} & 8 & 12 & 18 & $T_d$ & $z^{1}$ & 20.7 & 2.3 & 41 \\
\makecell[l]{\setlength{\tabcolsep}{0pt} Cuboctahedron} & 14 & 12 & 24 & $O_h$ & $x^{2}$ & 4.6 & 2.0 & 67 \\
\makecell[l]{\setlength{\tabcolsep}{0pt} Truncated \\~~ Cube} & 14 & 24 & 36 & $O_h$ & $x^{2}$ & 10.5 & 3.6 & 24 \\
\makecell[l]{\setlength{\tabcolsep}{0pt} Truncated \\~~ Octahedron} & 14 & 24 & 36 & $O_h$ & $x^{2}$ & 11.4 & 3.2 & 34 \\
\makecell[l]{\setlength{\tabcolsep}{0pt} Rhombicub- \\~~ octahedron} & 26 & 24 & 48 & $O_h$ & $x^{2}$ & 20.9 & 2.8 & 49 \\
\makecell[l]{\setlength{\tabcolsep}{0pt} Truncated \\~~ Cuboctahedron} & 26 & 48 & 72 & $O_h$ & $x^{2}$ & 24.5 & 4.6 & 21 \\
\makecell[l]{\setlength{\tabcolsep}{0pt} Snub cube} & 38 & 24 & 60 & $O$ & $x^{2}$ & 13.0 & 2.7 & 55 \\
\makecell[l]{\setlength{\tabcolsep}{0pt} Icosidodecahedron} & 32 & 30 & 60 & $\cellcolor{red!15} I_h$ & \cellcolor{red!15} $z^{4}$ & 5.7 & 3.2 & 39 \\
\makecell[l]{\setlength{\tabcolsep}{0pt} Truncated \\~~ Dodecahedron} & 32 & 60 & 90 & $\cellcolor{red!15} I_h$ & \cellcolor{red!15} $z^{4}$ & 12.9 & 5.9 & 13 \\
\rowcolor{yellow!40}\makecell[l]{\setlength{\tabcolsep}{0pt} Truncated \\~~ Icosahedron} & 32 & 60 & 90 & $\cellcolor{red!15} I_h$ & \cellcolor{red!15} $y^{4}$ & 14.3 & 5.0 & 22 \\
\makecell[l]{\setlength{\tabcolsep}{0pt} Rhombicosi- \\~~ dodecahedron} & 62 & 60 & 120 & $\cellcolor{red!15} I_h$ & \cellcolor{red!15} $z^{4}$ & 11.5 & 4.5 & 30 \\
\rowcolor{yellow!40}\makecell[l]{\setlength{\tabcolsep}{0pt} Truncated \\~~ Icosidodecahedron} & 62 & 120 & 180 & $\cellcolor{red!15} I_h$ & \cellcolor{red!15} $z^{4}$ & 19.8 & 7.6 & 12 \\
\makecell[l]{\setlength{\tabcolsep}{0pt} Snub dodecahedron} & 92 & 60 & 150 & $\cellcolor{red!15} I$ & \cellcolor{red!15} $z^{4}$ & 9.9 & 4.3 & 33 \\
\bottomrule
\end{tabular}
\end{table}

A Platonic solid can be assembled from magnetic spheres in contact such that the radius of the sphere is equal to half the length of the edge, $r_\mathrm{s} = a/2$. This construction is illustrated by the photograph of an octahedron shown in the inset of Fig.~\ref{fig:2}. The resulting central magnetic fields $B_\mathrm{c}$ for the five Platonic configurations are indicated by the red dots, with the corresponding exact values listed in Table~\ref{tab:solids}. The central fields of all polyhedra follow general rules, as detailed in Appendix~\ref{sec:appC} and~\ref{sec:appD}. Notably, all of these values exceed $B_\mathrm{c}/B_\mathrm{R} = 1/3$, which represents the maximum field attainable at the center of a hollow sphere embedded within an infinitely extended, homogeneously magnetized material.

It is instructive to compare these discrete results with estimates derived from continuum theory. To this end, the discrete values are rescaled by the volume ratio $v_\mathrm{r}$ between the total volume of $V$ magnetic spheres of radius $r_\mathrm{s} = a/2$ and that of the corresponding spherical shell, i.~e.,
\begin{equation}\label{eq:vol_ratio}
v_\mathrm{r} = \frac{V \cdot r_\mathrm{s}^3}{R_\mathrm{out}^3-R_\mathrm{in}^3}.
\end{equation}
The resulting continuum-based estimates are shown as open green circles. As expected, the deviation between these estimates and the exact discrete values decreases with increasing number of vertices $V$.

As a side note, applying the continuum-theory estimate via the volume fraction to the experimental value discussed in Appendix~\ref{app:finite ring} at least yields the correct order of magnitude: the measured field is underestimated by $12\%$, as indicated by the blue cross and open circle in Fig.~\ref{fig:2}.

\subsection{\label{sec:platonics} Magnetic fields of regular and semiregular polyhedral configurations}
Platonic and Archimedean polyhedra possess high symmetry and are therefore well suited for approximating the magnetization of continuum spherical shells. Although focused configurations can generate stronger magnetic fields, spherical Halbach configurations achieve perfect field homogeneity in the continuum limit \cite{Zijlstra1985}. Consequently, this arrangement is adopted in the following analysis.

\begin{figure}[ht]
\includegraphics[width=.48\textwidth]{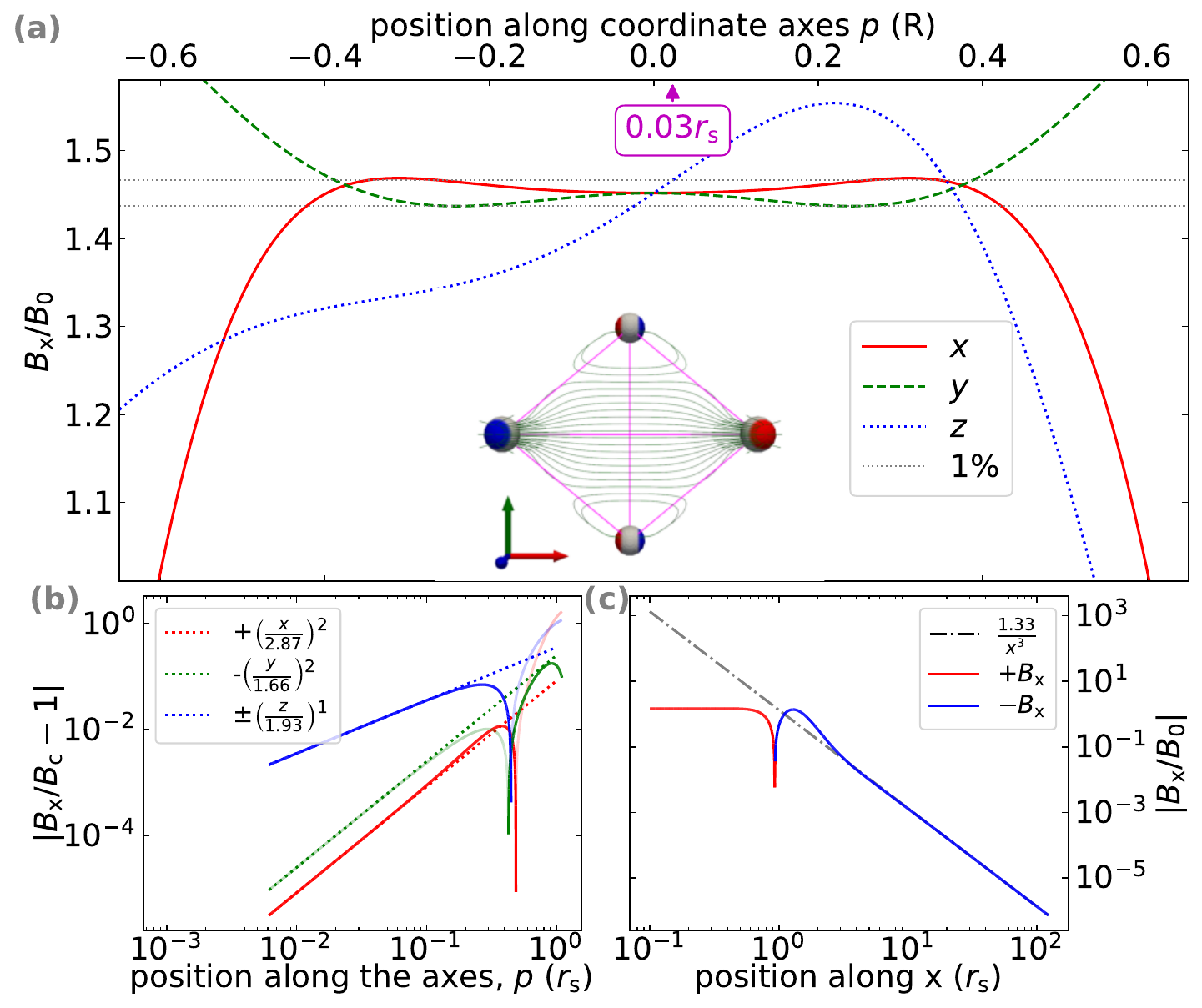}
\caption{Tetrahedral configuration: (a) Component $B_x$ along the $x$-axis (solid red), $y$-axis (dashed green), and $z$-axis (dotted blue). The dotted gray lines indicate $\pm1\,\%$ deviations from $B_\mathrm{c}$. The annotated number (magenta) on the upper $x$-axis gives the minimal radius where that deviation is reached. The inset shows the four point dipoles, represented as spheres (north pole in red, south pole in blue), together with magnetic field lines (solid green) in the $xy$-plane. The coordinate axes are color-coded to match the corresponding $B_x$ curves.
(b) Relative deviation $B_x/B_\mathrm{c}-1$ along the three coordinate axes, again shown in red, green, and blue. Sign changes are indicated by a transition from pale (negative) to darker (positive) shades. The dotted and dash-dotted straight lines represent fits to the asymptotic scaling, with the corresponding exponents and prefactors given in the panel.
(c) Magnetic field decay outside the cluster.}
\label{fig:Tetrahedron}
\end{figure}

The configuration with the smallest number of dipoles, corresponding to the four vertices of a tetrahedron, is shown in Fig.~\ref{fig:Tetrahedron}. The figure illustrates the magnetic field generated by this arrangement in the configuration defined by Eq.~(\ref{eq:Halbach}), with the field aligned along the $x$-axis.

The inset in Fig.~\ref{fig:Tetrahedron}(a) illustrates the positions and orientations of the four dipoles. Two magnets are located above the $xy$-plane, indicated by larger spheres, while the remaining two are positioned below this plane. This arrangement enforces mirror symmetry with respect to the planes $x=0$ and $y=0$. An additional mirror symmetry with respect to the $xy$-plane is not allowed within this symmetry group, since the remaining four symmetry axes of the tetrahedron are not perpendicular to these planes. The inset also includes representative magnetic field lines in the $z=0$ plane, which are shown only for qualitative guidance.

The spheres are used solely to visualize the orientations of the magnets, and their size does not have physical significance for point dipoles. Instead, point dipoles are fully characterized by their field strength $B_0$ at a distance $a/2$ along the dipole axis. The parameter $B_0$ provides a convenient scaling factor for the magnetic field profiles. Since the field of a point dipole can be realized by a uniformly magnetized sphere, it is worth noting that for spheres in contact, one has $B_0 = \tfrac{2}{3} B_\mathrm{R}$.

The $B_x(\rho)$ component of the magnetic field generated by the four dipoles is shown quantitatively in Fig.~\ref{fig:Tetrahedron}(a). Profiles along the three Cartesian directions, $\rho = x$, $\rho = y$, and $\rho = z$, demonstrate that the field remains symmetric with respect to the planes $x=0$ and $y=0$, since the Halbach condition in Eq.~(\ref{eq:Halbach}) preserves these symmetries. In contrast, the field profile $B_x(0,0,z)$, shown by the blue dotted line, does not exhibit such symmetry and crosses the center with a finite slope. The dashed gray lines indicate deviations of $\pm 1\,\%$ from the central field value, which are barely discernible on the linear scale.

The double-logarithmic plot in Fig.~\ref{fig:Tetrahedron}(b) enables an accurate characterization of the magnetic field and its spatial variation in the vicinity of the center. To this end, the relative deviation from the central field, $B_x(\rho)/B_\mathrm{c}-1$, is plotted. The ``curvature'' of this quantity along the principal axes is quantified in terms of characteristic length scales $\lambda$. They are defined as $B_x(\rho)/B_\mathrm{c}=1+(\rho/\lambda)^n$, where $n$ is the leading term of the Taylor expansion, which can be 1, 2, or 4 in the cases considered here. The values of $\lambda$ along the three Cartesian axes are listed in the figure caption. These lengths are normalized by $r_\mathrm{s}=a/2$, rather than by $R$, to allow a meaningful comparison between configurations with different radii $R$.

Interestingly, the tetrahedral arrangement in this orientation exhibits a second-order saddle point in the $xy$-plane, as reflected by the opposite signs of the curvatures reported in the caption. Specifically, the field displays a local minimum along the $x$-axis, as indicated by a positive curvature, and a local maximum along the $y$-axis, corresponding to a negative curvature. For larger values of $\rho$, the quantity $B_x/B_\mathrm{c}-1$ may change sign; this transition is visualized by a change in color intensity from pale (negative values) to dark (positive values).

Along the $z$-direction, the central field is neither a minimum nor a maximum. Instead, it exhibits a finite slope of $1/1.93\,r_\mathrm{s}^{-1}$. This imposes a severe limitation on the volume with a sufficiently constant field in this arrangement: A deviation from the center field smaller than $1\,\%$ is restricted to a height of $z=0.0193\, r_\mathrm{s}$.

To characterize the decay of the outer field, the quantity $|B_x(x,0,0)/B_0|$ is plotted, the color encoding the direction of $B_x$ (red for positive values and blue for negative values). The approximately homogeneous region of the field is clearly visible for $x<1$. At larger distances, the field exhibits an asymptotic $1/r^3$ decay, confirming that the configuration possesses a net dipole moment that is nonzero. The prefactor of 1.33 for this term corresponds to a truncated numerical value of $4/3$, which represents the net dipole moment of the configuration normalized to the magnetic moment of a single dipole. A finite dipole moment therefore renders this configuration qualitatively distinct from planar Halbach ring arrangements, which exhibit no external dipole moment.

\begin{figure}[ht]
\includegraphics[width=.48\textwidth]{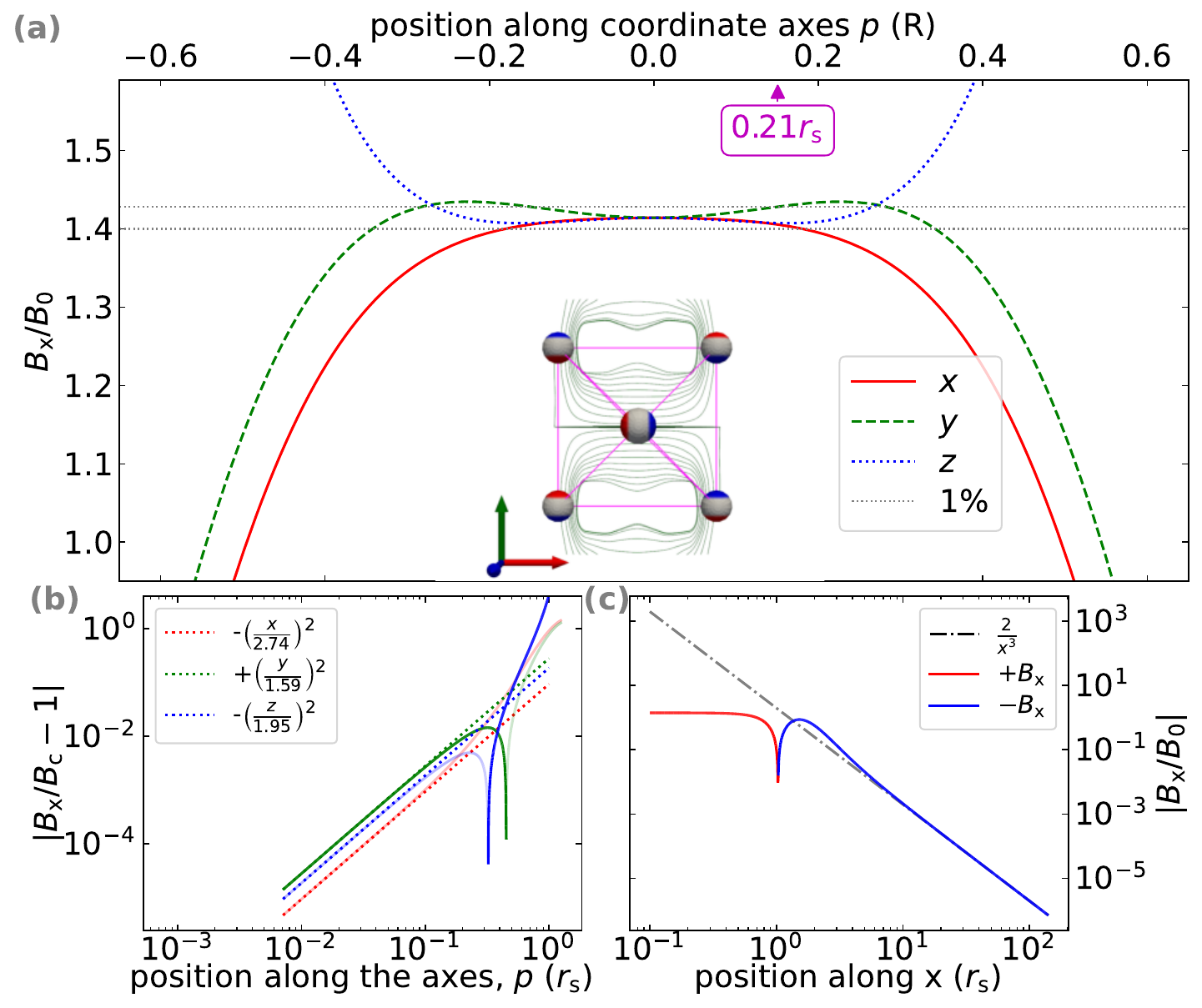}
\caption{Magnetic characteristics of the octahedral configuration. (a)–(c) follow the same layout and conventions as in Fig.~\ref{fig:Tetrahedron}.}
\label{fig:Octahedron}
\end{figure}

Although Fig.~\ref{fig:Tetrahedron} indicates that the tetrahedral arrangement has only very limited practical relevance due to its poor field homogeneity along the $z$-direction, the addition of only two magnets to form an octahedral configuration results in a substantial improvement, as shown in Fig.~\ref{fig:Octahedron}. In this geometry, four magnets are located in the $xy$-plane, with one magnet positioned above the plane and a corresponding one (not visible) below it.

Note that the dipole orientations are particularly simple in this case, being parallel or antiparallel to the $x$- and $y$-axes. The central field is flat to first order in all three spatial directions, since the octahedral symmetry class contains three mutually perpendicular mirror planes. Figure~\ref{fig:Octahedron}(b) illustrates several quantitative characteristics of this field. The characteristic length along the $x$-axis, $\lambda$=2.74, is smaller than that of the tetrahedral configuration, and the field strength, measured in units of $B_0$, is also slightly smaller. 

The net dipole moment is exactly $2m$ in this case, as indicated by the numerator of the 
$2/x^3$ term describing the asymptotic dashed–dotted gray line. The change in sign of the outer field is reflected by the color transition of the solid line: the field direction is positive near the center but becomes negative at larger distances from the octahedron.

\begin{figure}[ht]
\includegraphics[width=.48\textwidth]{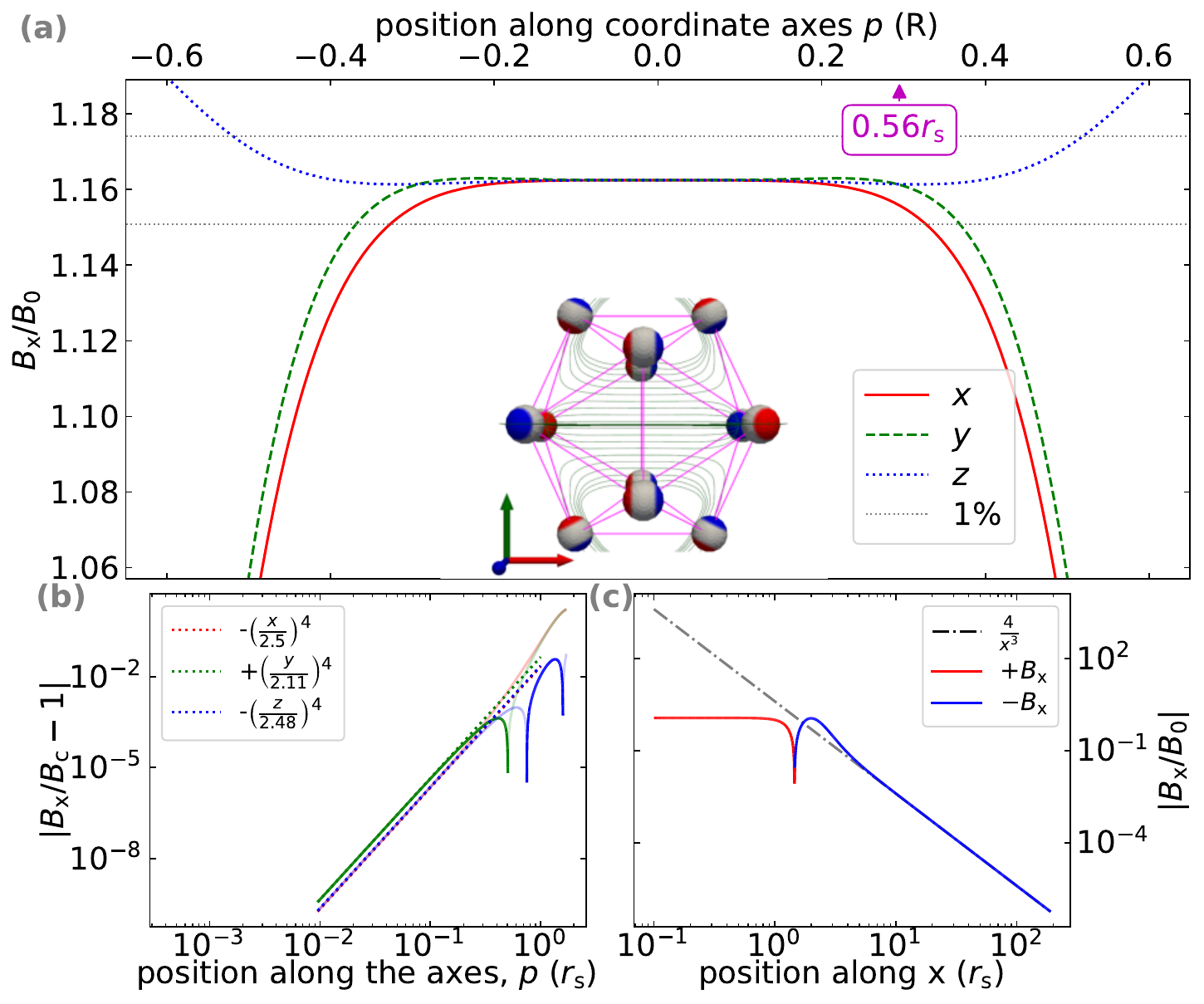}
\caption{Magnetic characteristics of the icosahedron configuration. (a), (b), and (c) follow the scheme of Fig.~\ref{fig:Tetrahedron}.}
\label{fig:Icosahedron}
\end{figure}

Using a larger number of spherical magnets in contact does not increase the maximum achievable field because the polyhedron must increase its radius to accommodate the additional magnets. However, this step can substantially improve the field homogeneity. This effect is illustrated in Fig.~\ref{fig:Icosahedron} for the icosahedral configuration with its 12 dipoles. In this case, the saddle point at the center acquires a new character: it is of fourth order, hence the field is varying as $\rho^4$ from the center. The ``steepest'' local variation occurs along the $y$-direction, which is not visible in Fig.~\ref{fig:Icosahedron}(a) but is clearly apparent in panel (b). A deviation of, for example, $1\,\%$ is reached at a distance $y=0.0211^{1/4} r_\mathrm{s}$.
Consequently, the volume within which the field remains within the $1\,\%$ deviation limit is  $(0.0211/0.0159^2)^{3/4}\approx 30$ times larger than for the octahedral arrangement.

\begin{figure}[ht]
\includegraphics[width=.48\textwidth]{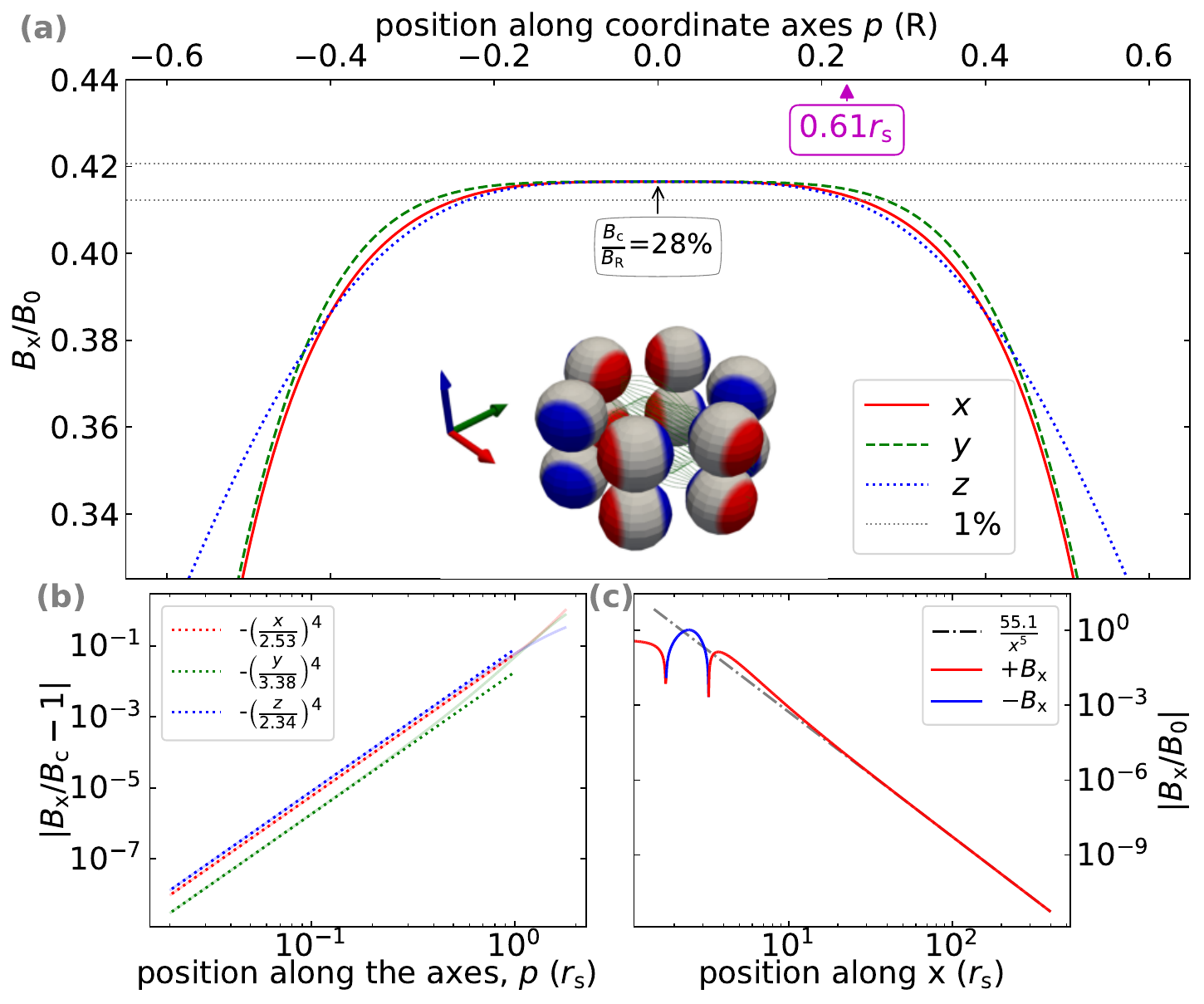}
\caption{Magnetic characteristics of a cylindrical Halbach sandwich at the optimal separation $R\sqrt{2/3}$. The sphere diameter is equal to the distance between the two rings. (a), (b), and (c) follow the scheme of Fig.~\ref{fig:Tetrahedron}.}
\label{fig:Sandwich}
\end{figure}

Before proceeding with the main line of argument, the following digression is worth recalling for context. Flat extrema and fourth-order saddle points can also be realized using Halbach rings in a cylindrical arrangement, provided that the separation between the two rings is chosen as  $R\sqrt{2/3}$. This well-known construction (Ref.~\cite{RehbergBlümler2025} and references therein) is shown in Fig.~\ref{fig:Sandwich}, where its magnetic field and homogeneity are compared with those of the icosahedral configurations described immediately above.

Two rings containing six spherical magnets each permit a direct comparison with the 12 vertices of the icosahedron. Figure~\ref{fig:Sandwich}(a) reveals a notable qualitative difference from Figs.~\ref{fig:Tetrahedron}–\ref{fig:Icosahedron}: the curvature is negative along all three shown directions. This does not imply the presence of a local maximum of the field; rather, the field still exhibits a saddle point, with directions of positive curvature located between the axes at angles $\varphi=60^\circ, 120^\circ, 240^\circ, \text{and }300^\circ$. Further details of this configuration can be explored interactively using open source software \cite{RehbergBlümler2025b}.

Another qualitative difference of the sandwich configuration compared with those discussed so far is evident in panel (c): the magnetic field decays with the fifth power of the distance, rather than with the third. This behavior reflects the fact that a single ring in a cylindrical Halbach orientation carries no net dipole moment, and the same property therefore holds for a sandwich composed of two such rings.

Comparing this optimized Halbach arrangement to the icosahedral cluster, the center field $B_\mathrm{c}$ of the latter is larger by a factor of $\beta=\frac{B_{\mathrm{c,iso}}}{B_{\mathrm{c,sw}}}\approx\frac{1.162}{0.416}\approx2.8$. This enhancement can be attributed to the closer packing of the magnets toward the center in the icosahedral arrangement.

A comparison of the homogeneity of the two arrangements reveals that the values of $\lambda$ are remarkably similar in both configurations. For a meaningful quantitative comparison, the fields must first be normalized to the same central value. The field produced by the optimized ring configuration cannot be further increased, since the spherical magnets are already in contact, as indicated in the inset of Fig.~\ref{fig:Sandwich}(a). In contrast, the field of the icosahedral configuration can be reduced by increasing the distance between the magnets by enlargement of the radius $R$. Setting the radius to $R\,\beta^{1/3}$ equalizes the central fields of the two configurations. This rescaling of $R$ by the factor $\beta^{1/3}$ increases each characteristic length $\lambda$ by the same factor, as the geometry remains similar. The ratio of volumes corresponding to a given level of field homogeneity can then be estimated from the three characteristic lengths as
$\frac{\mathrm{Vol_{iso}}}{\mathrm{Vol_{sw}}} \approx \frac{\beta\cdot2.5\cdot2.21\cdot2.48}{2.53\cdot3.38\cdot2.34}\approx 1.9$.

In the search for improved field homogeneity, all five Platonic solids and all thirteen Archimedean solids were investigated. The results are summarized in Table~\ref{tab:solids}. Each of these 18 configurations can be interactively explored with respect to their magnetic properties using an open source graphical user interface \cite{Rehberg2025b}. One key outcome is that the central saddle points cannot be flatter than fourth order for the convex uniform polyhedra discussed here. More importantly, icosahedral symmetry $I$ appears to be essential for achieving this highest order, a point that is further elucidated in Appendix~\ref{sec:appE}.

\begin{figure}[ht]
\includegraphics[width=.48\textwidth]{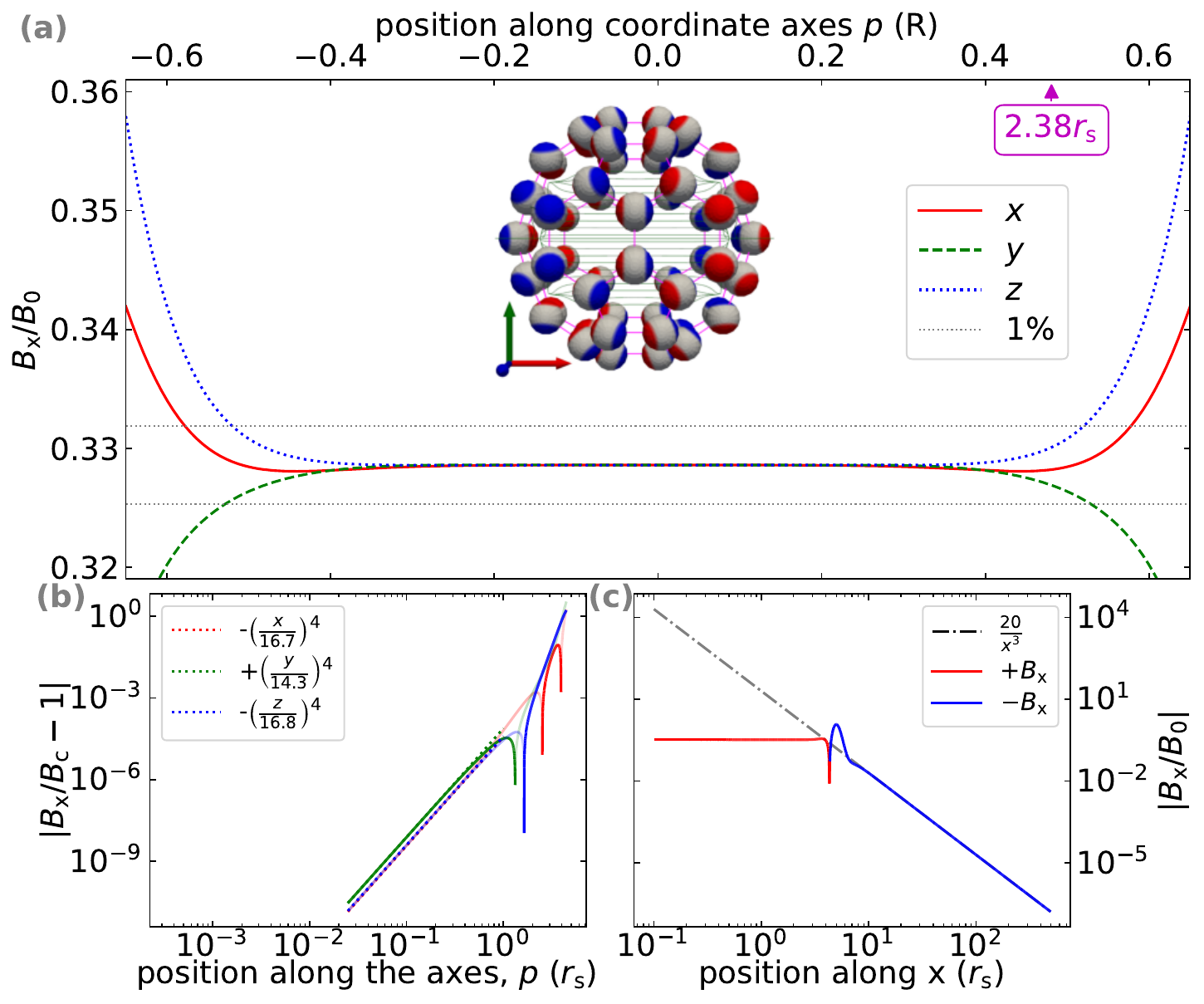}
\caption{Magnetic characteristics of the truncated icosahedron configuration.  (a), (b), and (c) follow the scheme of Fig.~\ref{fig:Tetrahedron}.}
\label{fig:Truncated icosahedron}
\end{figure}

Under this assumption, only six candidates from the set of Archimedean solids need to be considered. To identify the configuration with the best performance in terms of field homogeneity, the product of the three characteristic lengths is multiplied by the value of the central field, as in the comparison above. For a more stringent characterization, it is preferable to take the minimum of the three characteristic lengths, as listed in Table~\ref{tab:solids}, and raise it to the third power. This approach corresponds to comparing spherical volumes, which is of greater practical relevance than comparing ellipsoids of differing shapes. Consequently, the flux-like field-flatness product $f_\lambda=\lambda^3 B_\mathrm{c}$ is introduced. A comparison of this parameter between different configurations provides a precise and unambiguous measure of their homogeneous volumes, independent of whether homogeneity is specified on the ppm or percent scale. Using the values in Table~\ref{tab:solids}, the truncated icosidodecahedron and the truncated icosahedron emerge as the polyhedra of highest rank with respect to this metric. The latter has the familiar shape of a soccer ball, as shown in the inset of Fig.~\ref{fig:Truncated icosahedron}. This geometry is also known as the structure of the C$_{60}$ molecule, a fullerene commonly referred to as the “buckyball.”

This configuration involves 60 magnets. The resulting gain in field homogeneity is again substantial, even when compared with the icosahedral arrangement. The increase in usable volume for a given level of homogeneity is given by $(14.3/2.11)^3 \approx 300$. Note that the central field of the truncated icosahedron is approximately three times smaller, since the radius $R$ must be increased to accommodate the 60 magnets. When the field strength is taken into account by considering the ratio of the corresponding $f_\lambda$ values, the resulting gain in homogeneous field volume is reduced to a factor of about 125.

\begin{figure}[ht]
\includegraphics[width=.48\textwidth]{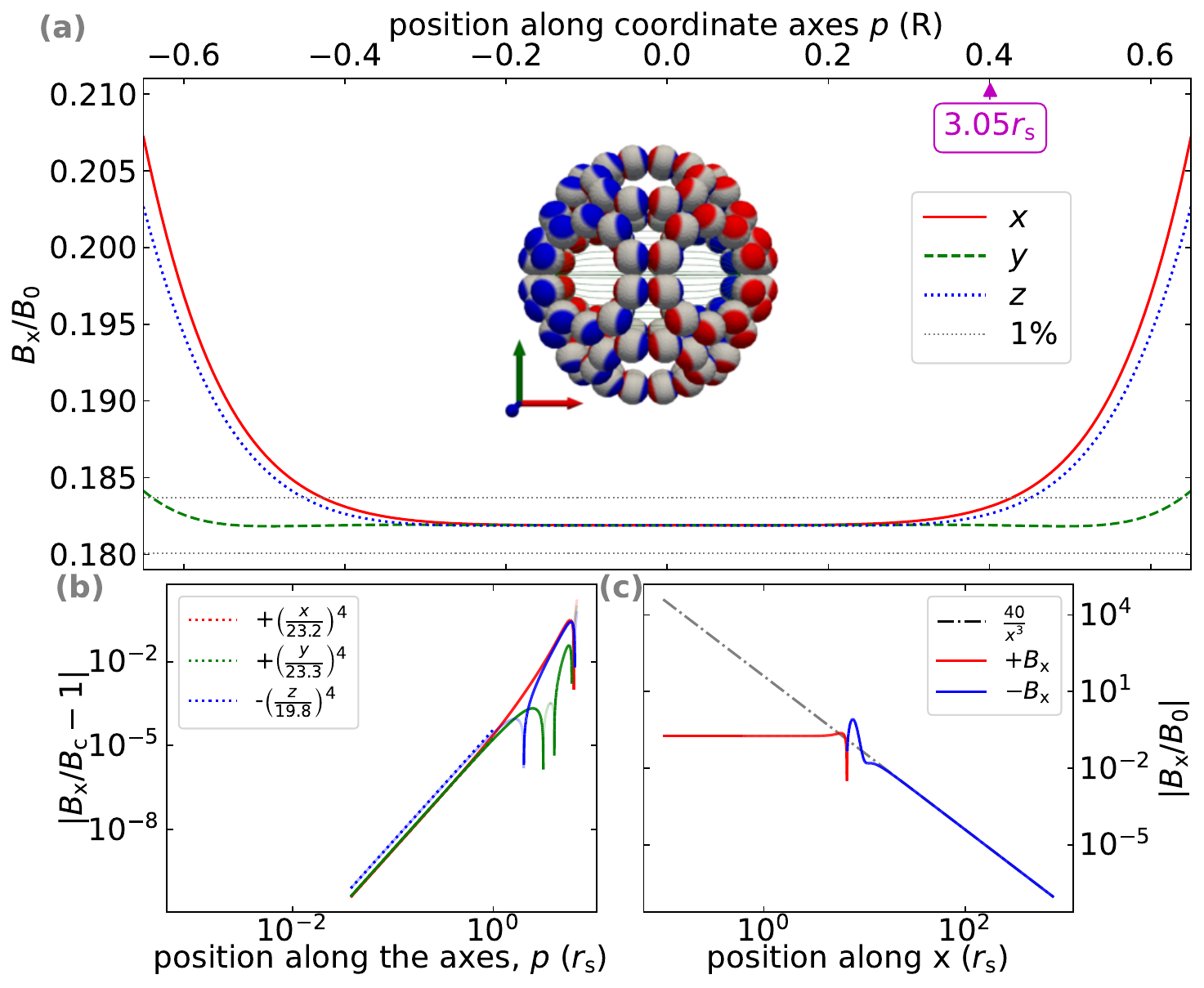}
\caption{Magnetic characteristics of the truncated icosidodeca\-hedron  configuration.   (a), (b), and (c) follow the scheme of Fig.~\ref{fig:Tetrahedron}.}
\label{fig:Truncated icosidodecahedron}
\end{figure}

Finally, it is tempting to examine the Archimedean solid with the largest number of magnets, namely the truncated icosidodecahedron with its 120 vertices. The resulting configuration is shown in Fig.~\ref{fig:Truncated icosidodecahedron}.

In this configuration, the field strength is further reduced, while homogeneity is increased, even relative to the truncated icosahedron. With respect to the parameter $f_\lambda$, this corresponds to a gain by a factor of 2. Compared with the sandwich configuration shown in Fig.~\ref{fig:Sandwich}, the overall gain factor is 260.\\
\indent A particularly useful feature of this Archimedean solid is highlighted in the inset, namely the presence of 12 decagons among the 62 faces forming the polyhedron. Even when the magnets attain their maximum size, $r_\mathrm{s}=a/2$, and are therefore in contact with their neighbors, the structure retains 12 large openings. These provide convenient access to observe the interior, a property that is crucial for many experimental applications.

\section{\label{sec:experimental}Experimental setup and results}
For the design, magnetic cubes were selected instead of spheres. Although spheres generate an exact point-dipole field \cite{Hartung2018}, which is advantageous for certain technical applications \cite{Hartung2021}, cubes offer superior practicality in positioning and alignment, as their edges provide an unambiguous reference for the magnetization direction. At the distances relevant to the present experiment, the resulting differences in field geometry are negligible and can be accurately accounted for \cite{Camacho2013, Bjork2023, Sosa2024}.

To demonstrate both the high field homogeneity and the practical feasibility of constructing such polyhedral magnet arrangements, three geometries with $I_h$ symmetry were selected: (a) a pair of icosahedra, (b) a truncated icosahedron, and (c) a truncated icosidodecahedron (see insets in Figs.~\ref{fig:exp3_4}, \ref{fig:exp3_5}, \ref{fig:exp3_6}, and \ref{fig:exp3_7}).

\begin{figure}[ht]
\includegraphics[width=.49\textwidth]{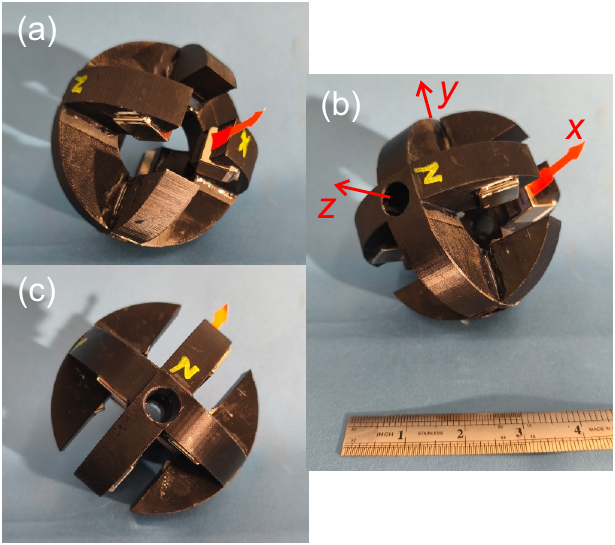}
\caption {Photographs of the magnet holder with the mounted magnets for the smaller icosahedron are shown from three different views. The direction of the magnetic field is along the $x$-axis, as indicated by a small red arrow temporarily attached to the holder. In panel (b), all coordinates are additionally indicated together with a scale for reference.}
\label{fig:exp1}
\end{figure}

The small icosahedron was realized from 12 cubical  Nd\textsubscript{2}Fe\textsubscript{14}B magnets ($20\!\times\!20\!\times\!20~\text{mm}^3$ of grade N45, item no. 3982 from Earthmag, Germany) positioned with their centers on a sphere of radius $R=41.85$~mm. The manufacturer specifies a remanence of $B_{\mathrm{R}} = 1.33-1.36$ T, while fitting the experimental data in Fig. \ref{fig:exp3_4} yields a value of  $B_{\mathrm{R}} = 1.37$~T. The magnets were bonded with epoxy into a PLA support structure fabricated by 3D-printing. The support structure is shown in Fig.~\ref{fig:exp1}. The assembled device  has a total mass of 915~g.
The large icosahedron was similarly equipped with 12 cubes measuring $30\!\times\!30\times\!30~\text{mm}^3$, each assembled from three $30\!\times\!30\times\!10~\text{mm}^3$ NdFeB blocks of grade N52 (item no.~QA-30x30x10-N52-N from maqna/otom, Germany). Their centers were placed on a sphere of radius $R=78.0$ mm. The specified remanence was $B_{\mathrm{R}} = 1.43-1.48$~T, and the experimentally determined value was 1.43~T. The magnets, together with the PLA support structure, amounted to a total mass of approximately 3~kg.

To construct the truncated icosahedron, 60 cubical magnets ($8\!\times\!8\!\times\!8~\text{mm}^3$; grade N45 item no. 2453, MagnetMax, Germany) were bonded onto a support using cyanoacrylate adhesive. The support was fabricated  from Tough2000 resin by stereolithography (Form 3 printer, Formlabs, Germany). The magnets were arranged so that their centers lay on a spherical surface of radius $R = 32$~mm. Although the manufacturer lists a remanence range of $B_{\mathrm{R}} = 1.32$–$1.38$~T, fitting the data in Fig. \ref{fig:exp3_6} resulted in a considerably lower effective value of $B_{\mathrm{R}} = 1.18$~T for this assembly. The completed device has a total mass of 260 g.

Finally, the truncated icosidodecahedron was assembled from 120 cubical magnets ($8\!\times\!8\times\!8~\text{mm}^3$, grade N50; item 2453 from Earthmag, Germany), arranged so that their centers lay on a spherical surface of radius $R = 50$~mm. The specified remanence range for these magnets was $B_{\mathrm{R}} = 1.41$–$1.43$~T, while experimentally only a value of $B_{\mathrm{R}} = 1.15$~T was determined. The support structure and the bonding method were manufactured identically to those used for the truncated icosahedron, and the final assembly had a total mass of 600~g.

All magnets were assembled manually. The 20~mm and 30~mm cubes were secured with C-clamps while the adhesive cured. The smaller 8~mm cubes were inserted into their sockets with a drop of cyanoacrylate and held in place for approximately 20~s using light manual pressure to ensure proper fixation.

The magnetic field generated by these magnets was measured using a 3-axis Hall probe (MV2 from Metrolab, Plan-les-Ouates, Switzerland), which is positioned in the $x$, $y$, and $z$ directions by three stepper motors using a custom-made positioning stage and software.

In the following, these field measurements for the four different configurations are presented. 

The first configuration is the small icosahedron constructed from 12 cuboids with a side length of $D_\mathrm{c}=20~\mathrm{mm}$ and arranged as shown in Fig.~\ref{fig:Icosahedron} on a radius $R=41.85~\mathrm{mm}$, which corresponds to the value used to program the 3D-printer. 

\begin{figure}[hbt]
\includegraphics[width=.48\textwidth]{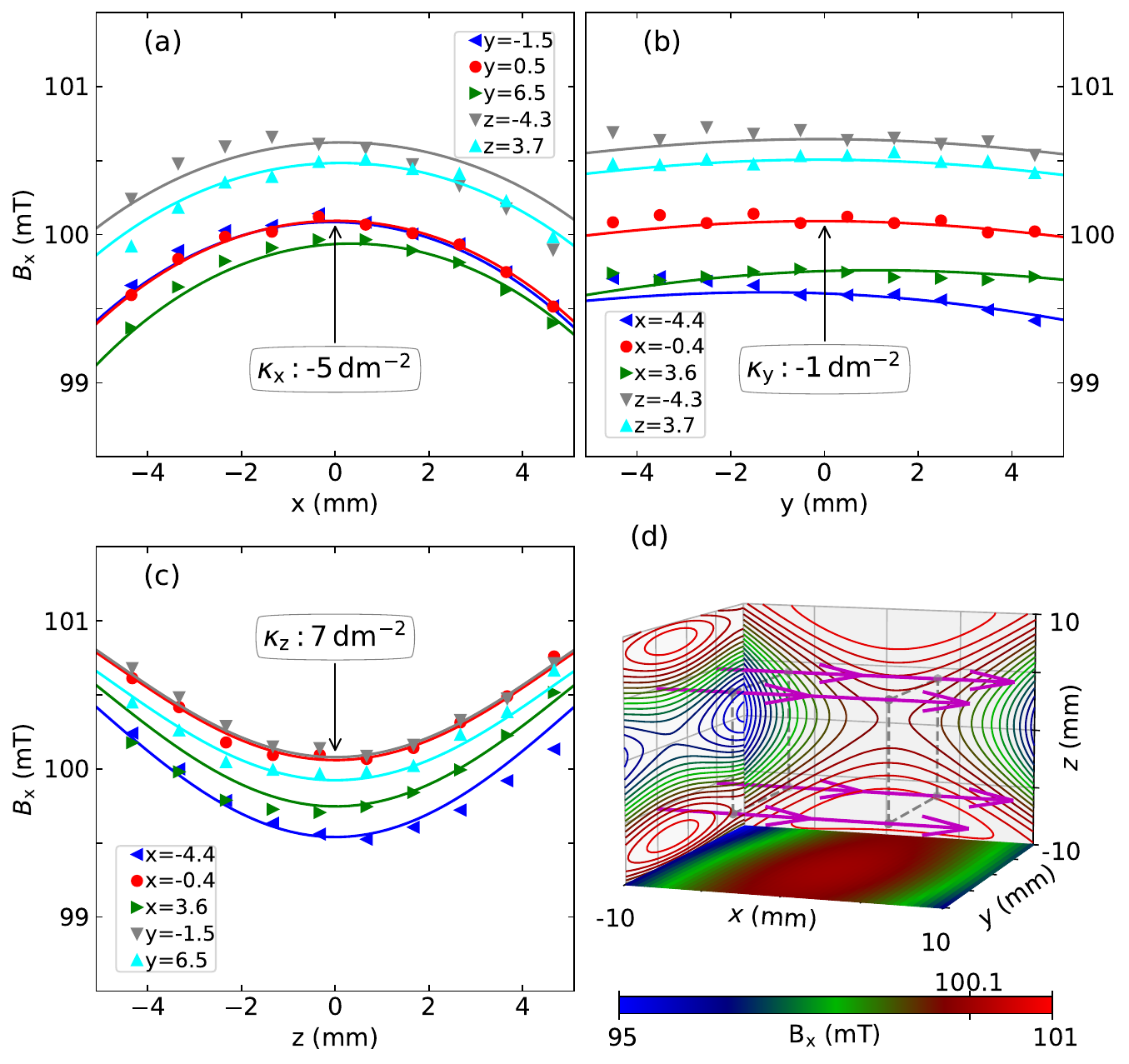}
\caption{Three dimensional measurements of $B_x$ taken near the center of the smaller icosahedron with $R=41.85~\mathrm{mm}$ and $D_\mathrm{c}=20~\mathrm{mm}$. (a), (b), and (c) show measurements along the three axes together with the corresponding fits, shown as solid lines. The insets give the curvature values. (d) provides a three dimensional representation of the measured field, shown via projections onto three planes.}
\label{fig:exp2}
\end{figure}

\begin{figure}[ht] 
\includegraphics[width=.48\textwidth]{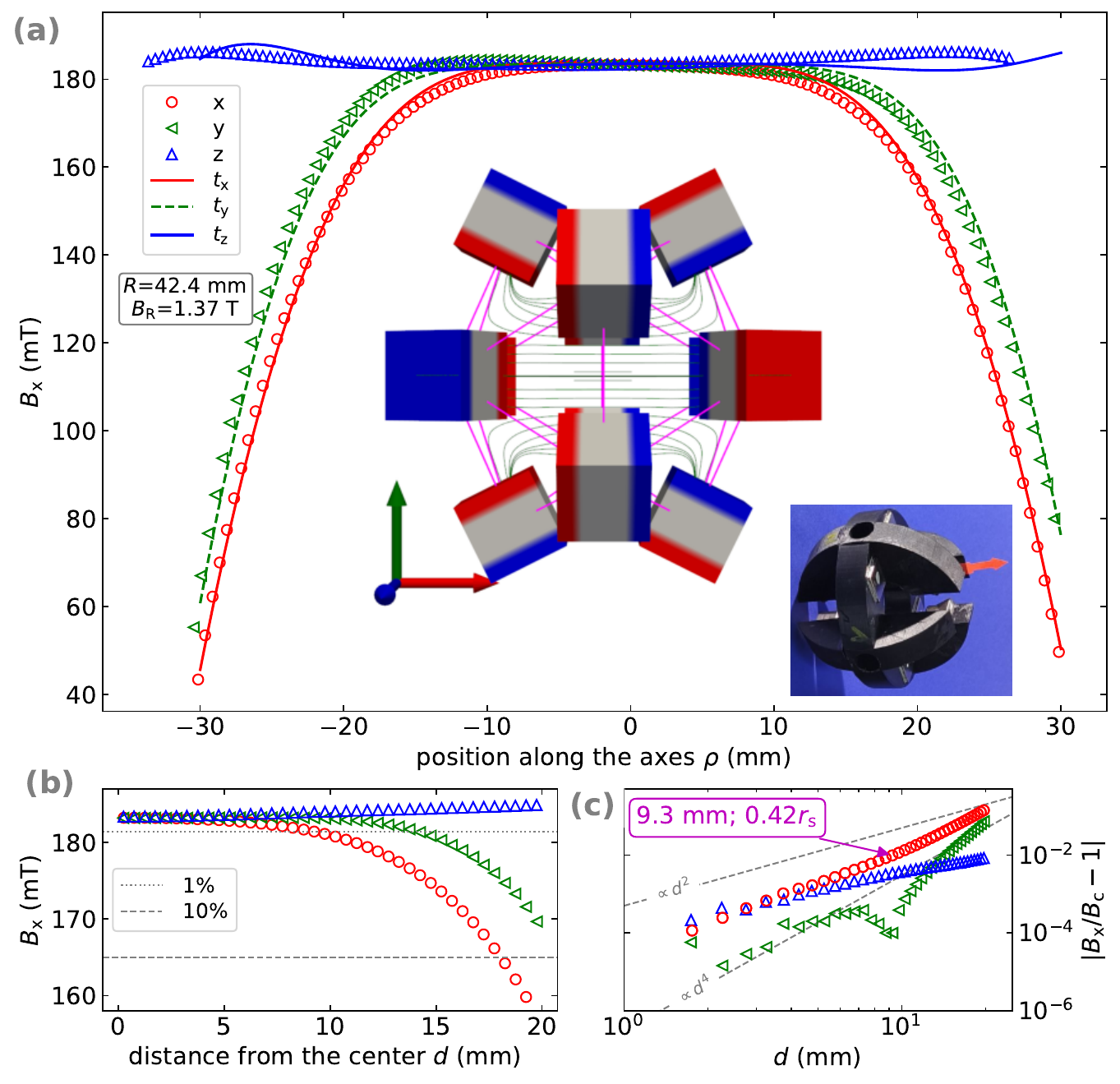}
\caption{Magnetic field measurements along the axes in the smaller icosahedron. The solid lines represent numerical solutions obtained for cubical magnets using the fit parameters $B_\mathrm{R}$ and $R$ (given in the box to the left). One inset shows the configuration constructed from $20~\mathrm{mm}$ cubes, viewed along the $xy$ plane; the edges of the icosahedron are indicated by magenta lines to guide the eye. The other inset shows a photograph of the magnet. (b) presents an enlarged view of the data near the center, with dashed lines indicating deviations of $1\,\%$ and $10\,\%$ from the central field value.  (c) shows the same data on a double-logarithmic scale, with dashed lines indicating slopes of 2 and 4, respectively.}
\label{fig:exp3_4}
\end{figure}

\begin{figure}[hbt]
\includegraphics[width=.48\textwidth]{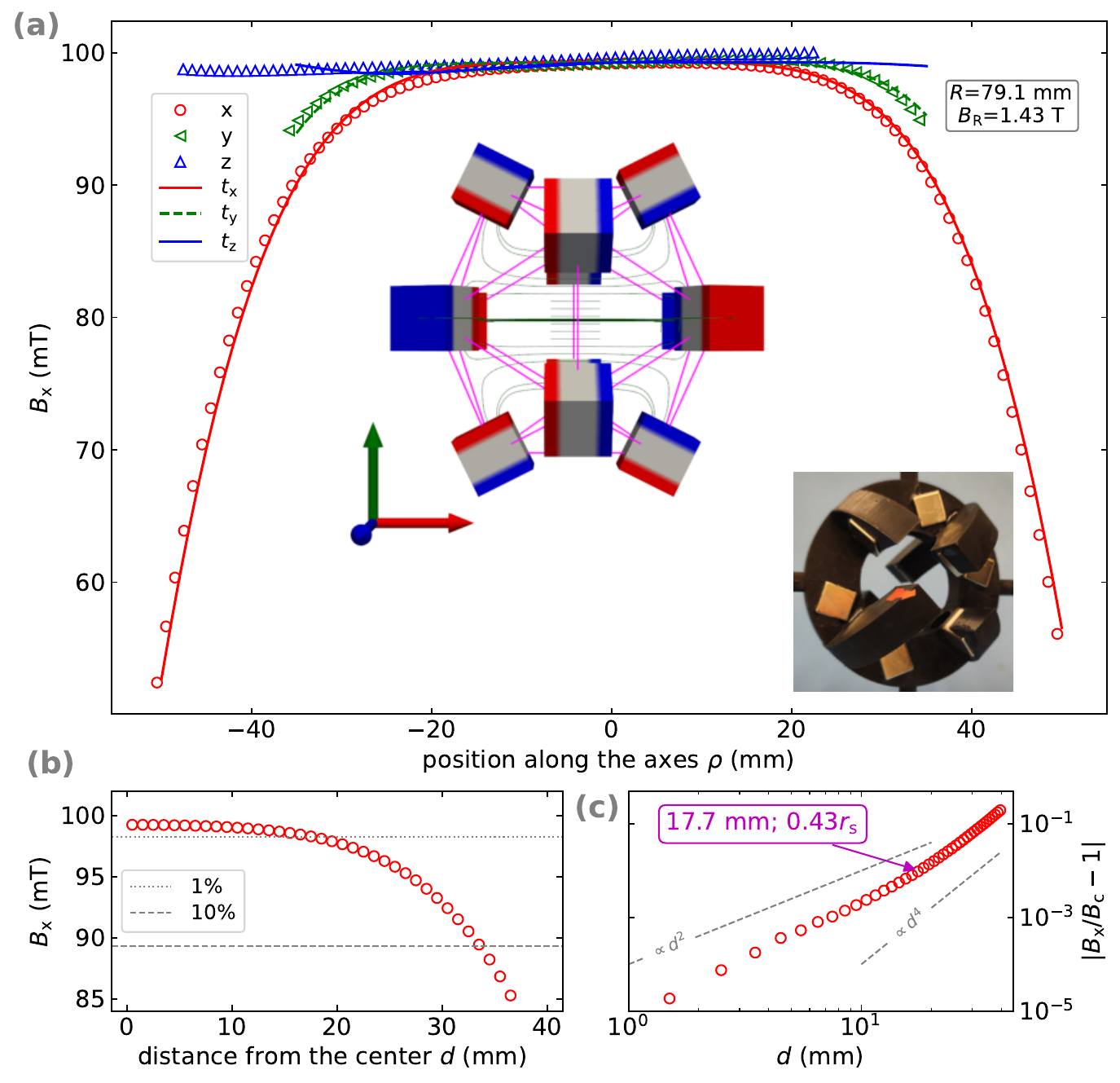}
\caption{Field measurements in the larger icosahedron. Same graphical representation and notation as in Fig.~\ref{fig:exp3_4}.} 
\label{fig:exp3_5}
\end{figure}

\begin{figure}[hb]
\includegraphics[width=.48\textwidth]{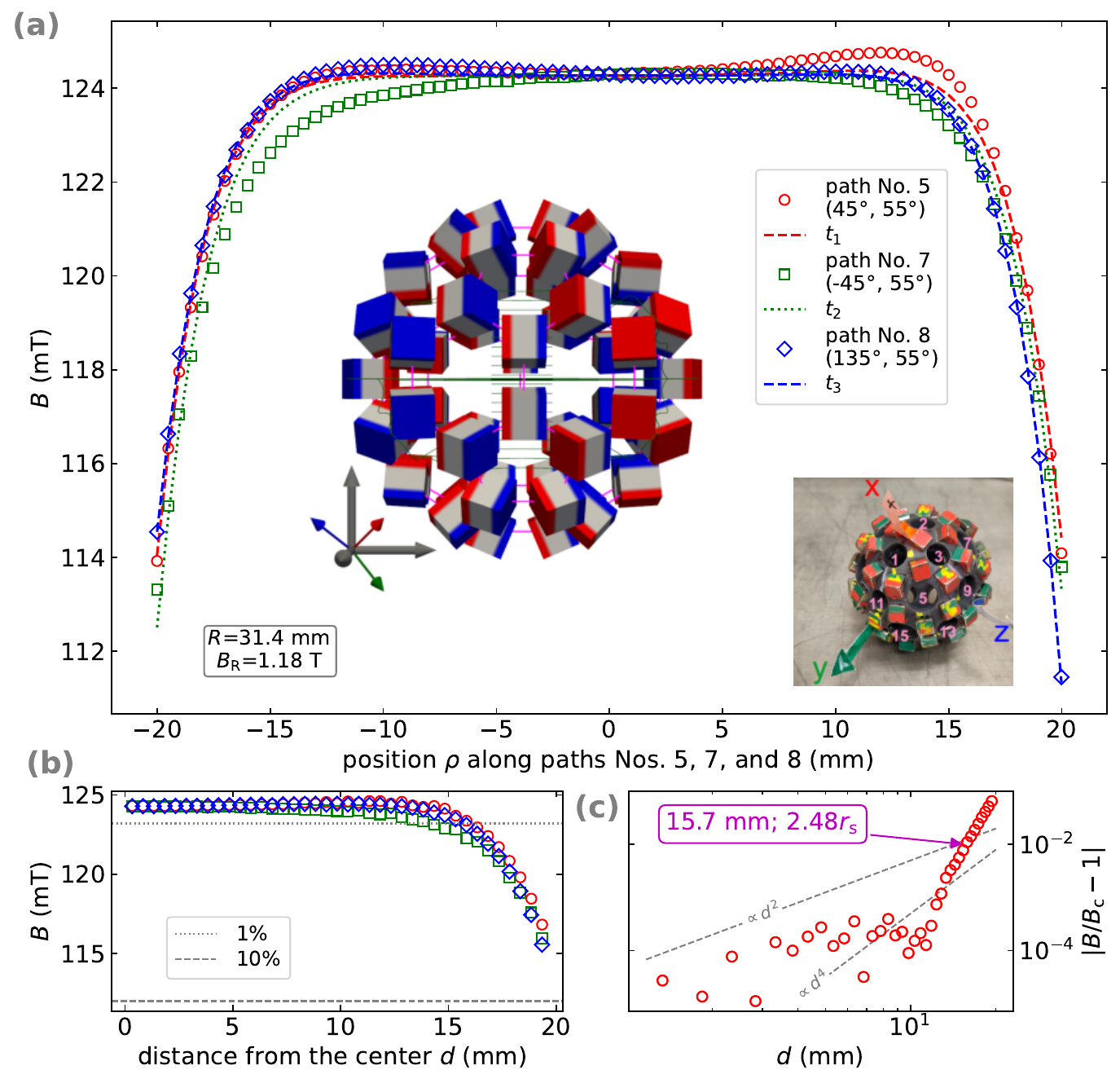}
\caption{Magnetic field measurements in the truncated icosahedron. Graphical representation and notation are similar to  Fig.~\ref{fig:exp3_4}. The path direction are given in the inset.}
\label{fig:exp3_6}
\end{figure}

\begin{figure}[ht] 
\includegraphics[width=.48\textwidth]{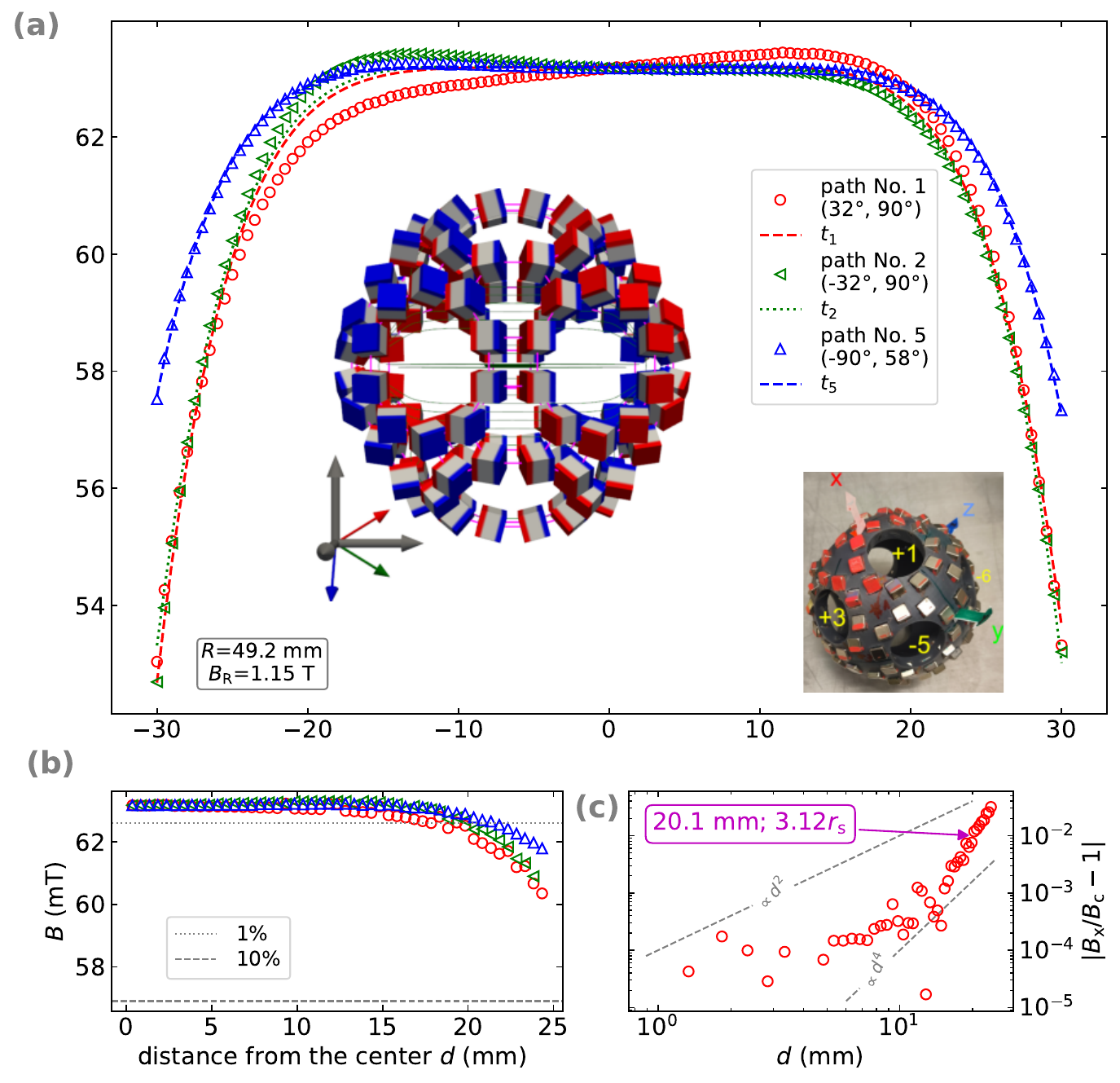}
\caption{Measurements in the truncated icosidodecahedron. The graphical representation and notation are similar to Fig.~\ref{fig:exp3_6}, but the path numbers indicate other directions here.}
\label{fig:exp3_7}
\end{figure}

The results of a three dimensional field measurement within a volume of $10~\mathrm{mm} \times 10~\mathrm{mm} \times 10~\mathrm{mm}$ are shown in Fig.~\ref{fig:exp2}. A total of 21 lines were measured along each of the three axes, resulting in $21\times21\times 21=9261$ measurement points for the three field components. In panel (a), five lines of the $B_x$ component measured along the $x$-axis, corresponding to the direction of the main magnetic field, are shown. 

In addition to the central line, two lines above and below, as well as two lines to the left and right, are displayed. This plot is used to define the $x$ coordinate of the center of the configuration. Panels (b) and (c) show analogous profiles measured along the $y$ and $z$ axes, respectively.
All lines are fitted with a Taylor-series ansatz that includes second- and fourth-order terms in the three coordinates. The fit provides a precise determination of the center position and the field curvatures along the three directions. The numerical values of the curvatures are given in the insets. They are defined by 
\begin{equation}
\frac{B_x}{B_\mathrm{c}} = 1 + \frac{\kappa}{2}\,\rho^{2} + \ldots
\end{equation}
i.~e., the characteristic length in this case is $\lambda=\sqrt{2/\kappa}$. Note that this quadratic term is not expected on the basis of idealized theory. This arises from our use of magnetic cubes, which are easier to orient than spheres, and from the unavoidable symmetry breaking introduced by this choice in the experimental setup. An illustration of the three dimensional fitting function is shown in panel (d). The magnetic field strength is represented by a color map in the $z=-10~\mathrm{mm}$ plane and by contour lines in the $x=-10~\mathrm{mm}$ and $y=-10~\mathrm{mm}$ planes. The eight arrows indicate the measured field direction at the corners of a $10~\mathrm{mm}$ cube. The most prominent qualitative feature of the central saddle point \textendash{} namely, a field maximum with respect to $x$ and $y$ and a minimum with respect to $z$ \textendash{} is consistent with the idealized point-dipole theory shown in Fig.~\ref{fig:Icosahedron}.

To obtain an overview on a larger spatial scale, such three dimensional measurements are not practicable due to the geometric constraints imposed by the magnet holder frame. Instead, extended scans can be performed along each of the three axes individually. The corresponding results on this expanded length scale are shown in Fig.~\ref{fig:exp3_4}. The theoretically expected values are indicated by dashed and dotted lines for comparison. Both $R$ and $B_\mathrm{R}$ are determined as common fit parameters for all measurement points.

The inset illustrates the configuration of the 12 cubical magnets, with red and blue indicating the north and south poles, respectively. In addition, a photograph of the magnet is shown.

Figure~\ref{fig:exp3_4}(a) provides an overview of the magnetic field within the frame. To more clearly quantify the degree of homogeneity near the center, Fig.~\ref{fig:exp3_4}(b) shows the field as a function of the distance from the center on a linear scale. This representation demonstrates that the field deviates from the central value $B_\mathrm{c}$ by less than $1\,\%$ within an approximately $1~\mathrm{cm}^3$ cubical volume. 

Figure~\ref{fig:exp3_4}(c) shows the same data on a logarithmic scale. The initial slope is closer to a second-order dependence than to the fourth-order behavior predicted by theory. This indicates that the fourth-order term represents a singular ideal case that is readily degraded by even small imperfections that break the ideal symmetry of the experimental setup. A crossover toward fourth-order scaling becomes apparent at distances of approximately 10~mm. Despite this limitation, the practical advantage of the present construction remains clear: a field homogeneity better than $5\,\%$ is achieved within a region of $20~\mathrm{mm}$ diameter around the center (within the inner accessible volume of approximately $30~\mathrm{mm}$ radius).

Applications of this simplest configuration with icosahedral symmetry are not restricted to the size to radius ratio used in Fig.~\ref{fig:exp3_4}, nor to the field strength obtained there. A practicable example is illustrated by the wish to create a homogeneous field near $100~\mathrm{mT}$, and by maximizing the volume of the homogeneous field. This implies to use the largest possible cubes. Due to our experience, the upper boundary for handling these strong magnets is around an edge length 3~cm. Starting with this constraint, one increases the radius up to the value where the desired field is reached, thus a larger icosahedron was constructed with radius $R=78~\mathrm{mm}$ and cuboids of edge length $D_\mathrm{c}=30~\mathrm{mm}$, as shown in Fig.~\ref{fig:exp3_5}.\\

The analysis focuses on measurements along the $x$-axis, which appears to be the direction that exhibits the poorest field homogeneity. It appears to be dominated by quadratic terms up to about 1~cm. A deviation of 1\% is reached at a distance of about 2~cm. This is below the value expected from the idealized theory presented in Fig.~5, which yields $0.32 R \approx 25\,\mathrm{mm}$. These deviations reflect the imperfections of our geometrical assembly and the fact that the individual magnets differ on percentage level. They are not caused by the fact that here we are comparing the theory for spheres rather than for cubes. The deviations between both magnet types would ideally be smaller, as can be explored interactively with the software provided in \cite{RehbergBlümler2025b}.

A higher degree of field homogeneity is expected for Archimedean configurations with a larger number of magnets within the same symmetry class. A truncated icosahedron was constructed from 60 smaller cubical magnets with a side length of $D_\mathrm{c}=8~\mathrm{mm}$. A photograph is shown in the inset of Fig.~\ref{fig:exp3_6}(a).

The structure is oriented to preserve symmetry with respect to the three Cartesian planes. In this configuration, measurements along the $x$, $y$, and $z$ axes are not feasible due to the dense packing of the magnets. Instead, measurement access is provided by openings centered on the 32 faces of the truncated icosahedron. The 12 regular pentagonal faces and 20 regular hexagonal faces allow for 16 measurement paths: six connecting opposite pentagons and ten connecting opposite hexagons. These paths are numbered for reference and their directions are specified in spherical coordinates; for example, path No.~5 corresponds to the polar angle $\theta=45^\circ$ and the azimuthal angle $\phi=55^\circ$, as indicated in parentheses in the legend. In addition, three representative paths are indicated by red, green, and blue arrows, together with the Cartesian coordinate axes shown in gray.

Along these different paths, the three Hall sensors inside the 3D probe are oriented at oblique angles with respect to the Cartesian coordinate axes. In principle, the component $B_x$ could be obtained from an appropriately weighted combination of their signals. However, this approach would introduce an additional source of uncertainty due to the finite precision of the sensor orientations. This issue is avoided by using the field magnitude $B=\sqrt{B_x^2+B_y^2+B_z^2}$ instead. This quantity is independent of the orientation of the probe and coincides with the central value of $B_x$. Therefore, it is shown on the vertical axis of Fig.~\ref{fig:exp3_6}.

The measurements reveal systematic asymmetries along paths Nos.~5 and~7. These effects are at the sub-percent level and are presumably caused by small imperfections in the fabrication of the assembly. Possible sources include unavoidable variations in the magnetic properties of the 60 magnets, for which deviations at the percent level are common, as well as small inaccuracies in positioning the relatively small magnets within the support. In addition, minor misalignments of the measurement paths may be present, since an angular precision better than $\pm2^\circ$ is difficult to achieve.

Despite these uncertainties, the enlarged view in Fig.~\ref{fig:exp3_6}(b) demonstrates that a working volume of $25~\mathrm{mm} \times 25~\mathrm{mm} \times 25~\mathrm{mm}$ with a field deviation of less than $1\,\%$ can be realized with this apparatus. This performance can be compared with the working volume expected theoretically for point dipoles: according to Fig.~\ref{fig:Truncated icosahedron}, the diameter corresponding to a $1\,\%$ deviation is $2 \cdot 0.48 R$, which yields $30~\mathrm{mm}$ for a radius of $R=31~\mathrm{mm}$. This deviation is not explained by the fact that cuboids instead of spheres are used, because both magnets are expected to show rather similar performance at this 1\% level \cite{RehbergBlümler2025b}. The discrepancy is explained rather by the experimental uncertainties listed above.

Note that the remanence $B_{\text{R}}$ of the 8~mm cubes seems to be lower than that obtained for the larger cubes. A possible explanation could be the larger influence of the non-magnetized Ni/Cu/Ni coating layer, but this issue is not investigated in further detail here. 

The highest level of field homogeneity within the icosahedral class is expected for the truncated icosidodecahedron, as indicated in Table~\ref{tab:solids}. Experimentally, this configuration presents the greatest challenge due to the requirement to position 120 magnets. The corresponding results are shown in Fig.~\ref{fig:exp3_7}. As in the case of the truncated icosahedron, direct measurements along the three Cartesian axes are not possible. Instead, six convenient measurement paths are available that connect the centers of the 12 regular decagonal faces. These paths are numbered for reference; however, the numbering does not correspond to that used for the truncated icosahedron.

The maximum field of approximately $63~\mathrm{mT}$ is well below the theoretical maximum of $0.12\,B_\mathrm{R}$, because the nonmagnetic material of the magnet holder now constitutes a substantial fraction of the shell volume. The field homogeneity is improved compared to that of the truncated icosahedron: the volume within which the field deviates by less than $1\,\%$ is approximately $30~\mathrm{mm} \times 30~\mathrm{mm} \times 30~\mathrm{mm}$. Moreover, the large observation windows provided by the decagonal faces (here 32~mm in diameter) represent a major experimental advantage of this geometry.

\section{Conclusion and Outlook}
In the pursuit of experimentally accessible Halbach spheres, distributions of permanent magnets on spherical shells were investigated as a natural route toward approximating the continuously varying ideal magnetization. Regular (Platonic) and semiregular (Archimedean) polyhedral arrangements offer two distinct advantages: all magnets lie at the same distance from the center, and each magnet has a unique nearest-neighbor distance. The first property ensures that one gets the maximal field contribution from every single magnet within the restriction by that minimal distance. The second advantage is more of a practical nature: One defines the minimal thickness between the holes in the supporting 3D-printed spherical shell once and for all, thus optimizing the material usage. Moreover, these polyhedral arrangements have the practical advantage of being constructed from identical parts.

These polyhedra fall into three symmetry groups. An analysis of the magnetic-field homogeneity produced by dipoles in Halbach orientation disqualifies the tetrahedral class and identifies icosahedral symmetry as the most favorable, yielding a fourth-order saddle point of the scalar magnetic potential at the center. This conclusion follows from the structure of solutions to the Laplace equation under the symmetry constraints of the relevant point groups, with key steps summarized in Appendix~\ref{sec:appE}.

Prototype Halbach spheres were constructed from discrete cubical magnets in four different configurations with icosahedral symmetry. Measurements confirm that the magnetic field near the center is remarkably homogeneous, even without additional shimming methods. Although spherical shells naturally restrict access to their interiors, polyhedral discretization partially mitigates this challenge, because the shells inherently contain large openings. In the Archimedean design comprising 120 magnets (the truncated icosidodecahedron), the largest apertures form twelve regular decagons, providing substantial access to the interior volume.

Beyond offering improved field homogeneity over a large fraction of the sample space, such polyhedral magnet assemblies can serve as compact and inexpensive substitutes for vector (three-axis) magnets in experiments requiring strong and homogeneous fields at tunable orientations (e.g. in many condensed-matter studies~\cite{Galvis2015} and quantum-materials studies~\cite{Mrozek2015}). Field-amplitude control could be achieved by employing two concentric polyhedra that rotate relative to each other around an axis perpendicular to the generated field (analogous to~\cite{Leupold1988,Soltner2023}). However, unlike cylindrical arrangements of Halbach rings, the dipole moment of the spherical arrangements discussed here is different from zero. Hence, there will be a torque when rotating them perpendicular to the field axis, whose influence would need to be studied.

Another potentially promising application is magnet rotation at the “magic angle” with respect to a static sample in magnetic resonance experiments, allowing the averaging of weak dipolar couplings without mechanically rotating the sample itself~\cite{Meriles2004,Wind2006,Sakellariou2010}.

As a direction for future work, concepts analogous to the force-free opening angles identified in ~\cite{patent_spheres,BluCasa2015} may be applicable to these polyhedral shells, potentially enabling easily accessible, reconfigurable, or adjustable Halbach sphere architectures.

Additionally, Halbach hemispheres can be combined with Halbach cylinders as their end caps (i.~e., spherocylinder or capsule) \cite{Soltner2023, Chen2007}. Together with the concept of force-free opening, this approach enables the construction of extended assemblies that maintain a highly homogeneous magnetic field over the entire inner volume.

Together, these results open a path toward the practical realization of Halbach’s spherical ideal: a permanent-magnet architecture that unites geometric elegance with exceptional field quality. The constructions demonstrated here constitute the first practical implementations of discretized Halbach spheres and illustrate their superior homogeneity. This work may provide a foundation for next-generation permanent-magnet assemblies, enabling unprecedented field uniformity and offering geometric configurations ripe for further exploration.

\begin{acknowledgments}
 We thank Andreas Engel for helpful suggestions. Lukas Schmidt is acknowledged for help with stereolithographic printing of the magnet supports. 
\end{acknowledgments}

\appendix
\section{Calculation of the center field}
\label{app:continua}
The magnetic field at the center, obtained from Eq.~(\ref{eq:continua}) and expressed in spherical coordinates, reads
\begin{equation*}
\mathbf{B}(\mathbf{0}) = \frac{\mu_0}{4\pi} \int_{\mathrm{Vol}}  \frac{3 (\mathbf{M}(\mathbf{r}) \cdot \hat{\mathbf{r}}) \hat{\mathbf{r}} - \mathbf{M}(\mathbf{r})}{|\mathbf{r}|^3}  \, d^3r
\end{equation*}
\begin{equation}
= \frac{\mu_0}{4\pi} \int_{R_\mathrm{in}}^{R_\mathrm{out}} \int_0^\pi \int_0^{2\pi}\frac{3 (\mathbf{M} \cdot \hat{\mathbf{r}}) \hat{\mathbf{r}} - \mathbf{M}}{r^3} r^2 \sin\theta \, d\phi\, d\theta \, dr.
\end{equation}
For a rotationally symmetric magnetization distribution with constant remanence $B_\mathrm{R}$
\begin{equation}
\mathbf{M} = \frac{B_\mathrm{R}}{\mu_0} \begin{pmatrix}
\sin(\alpha) \cos(\phi) \\
\sin(\alpha) \sin(\phi) \\
\cos(\alpha)
\end{pmatrix}, 
\end{equation}
and the integration over $\phi$ contributing a factor of $2\pi$, this leads to the magnitude of the magnetic field
\begin{equation*}
B_\mathrm{c} = \frac{B_\mathrm{R}}{2} 
\int_{R_\mathrm{in}}^{R_\mathrm{out}} 
\int_0^\pi \left[ 
3 \sin(\alpha) \sin\theta \cos\theta  
 \right. 
\end{equation*}
\begin{equation}
\left. + 3 \cos(\alpha) \cos^2\theta - \cos(\alpha) \right] \sin\theta \, d\theta \,\frac{1}{r} dr.
\end{equation}
For the Halbach configuration defined by Eq.~(\ref{eq:Halbach}), this integral yields
\begin{equation}
B_\mathrm{c} = \frac{4}{3} B_\mathrm{R}  \cdot \ln\left( \frac{R_{\mathrm{out}}}{R_{\mathrm{in}}} \right),
\end{equation}
while for the focused configuration (Eq.~(\ref{eq:focused})) the result is
\begin{equation}
B_\mathrm{c} = \left[1+\frac{\ln\left(2+\sqrt{3}\right)}{2\sqrt{3}}\right] B_\mathrm{R}  \cdot \ln\left( \frac{R_{\mathrm{out}}} {R_{\mathrm{in}}} \right).
\end{equation}
For a thin annular disk of thickness $h \ll R_\mathrm{in}$, the volume integral reduces
to a surface integral,
\begin{equation}
\begin{split}
B_\mathrm{c} = \frac{B_r h}{4\pi} 
\int_{R_\mathrm{in}}^{R_\mathrm{out}} 
\int_0^{2\pi} \bigl[
3 \sin(\alpha) \sin\theta \cos\theta
\\
+\, 3 \cos(\alpha) \cos^2\theta
- \cos(\alpha)
\bigr]
\, d\theta \,\frac{1}{r^2}\,dr.
\end{split}
\end{equation}
In the Halbach case (Eq.~(\ref{eq:Halbach})), this yields
\begin{equation}
B_\mathrm{c} = \frac{3}{4} B_\mathrm{R}  \cdot \left[ \frac{h} {R_{\mathrm{in}}}-\frac{h} {R_{\mathrm{out}}} \right],
\end{equation}
while for the focused configuration (Eq.~(\ref{eq:focused})) the result is
\begin{equation}
B_\mathrm{c} = \frac{2}{\pi}\text{E}\left(\frac{3}{4}\right) B_\mathrm{R}  \cdot \left[ \frac{h} {R_{\mathrm{in}}}-\frac{h} {R_{\mathrm{out}}} \right].
\end{equation}

\section{\label{app:finite ring} Applying continuum theory to discrete assemblies}
The theory developed for continuous disks and illustrated in Fig.~\ref{fig:1} can also be employed to estimate the magnetic field at the center of a finite ring assembled from discrete magnets. As an example, the difference between the values at $R_\mathrm{in}/h = 3$ and $R_\mathrm{out}/h = 4$ corresponds to a central field of $B_\mathrm{c} \approx 0.064 B_\mathrm{R}$, as indicated by the separation between the dashed-dotted lines in the figure. This approach can be applied to the experimental configuration shown in the inset, consisting of 16 cubical magnets of volume $1~\text{cm}^3$ arranged on a ring with a radius of $R = 35~\text{mm}$, i.~e., with $R_\mathrm{in} = 30~\text{mm}$ and $R_\mathrm{out} = 40~\text{mm}$, as reported in Ref.~\cite{RehbergBlümler2025}. The resulting central field can be estimated by scaling the value obtained from Eq.~(\ref{eq:fBc}) by the fraction of the ring area actually covered by the magnets,
\begin{equation}
\frac{16 \cdot a_\mathrm{c}}{(R_\mathrm{out}^2-R_\mathrm{in}^2)\pi}.
\label{eq:5}
\end{equation}
For $a_\mathrm{c}\!=\!1~\text{cm}^2$ and $B_\mathrm{R}=1.33~\text{T}$, this results in a central field of $B_\mathrm{c}=62.17~\text{mT}$, which agrees with the measured value with a deviation less than 0.5\%~\cite{RehbergBlümler2025}. This approximation is also available in an interactive format through a graphical user interface (GUI) provided as open source software \cite{RehbergBlümler2025b}.

\section{\label{sec:appC} The center field of polyhedral clusters}
The analog of Eq.~(\ref{eq:continua}) for a regular or semiregular polyhedron of point dipoles located at the $V$ vertices in Halbach orientation as in Eq. (\ref{eq:Halbach}) reads
\begin{equation}\label{eq:C1}
\begin{aligned}
B_\mathrm{c} &= \frac{B_\mathrm{0}}{2} \left(\frac{r_s}{R}\right)^3 
\sum_{i=1}^V \Bigl[
  3 \sin(2\theta_i) \sin\theta_i \cos\theta_i \\
&\quad + 3 \cos(2\theta_i) \cos^2\theta_i
  - \cos(2\theta_i)
\Bigr]\\
&= \frac{B_\mathrm{0}}{4} \left(\frac{r_s}{R}\right)^3 
\sum_{i=1}^V \Bigl[
  \cos(2\theta_i) +3 \Bigr],
\end{aligned}
\end{equation}
where $\theta_i$ is the angle between the $i$\textsuperscript{th} dipole moment and the field orientation, which is the $x$-axis in this paper. 
This term reveals that dipoles located on the $x$-axis contribute most strongly to the field. Evaluating this sum —using computer algebra— for all 18 regular and semiregular polyhedra yields the same result for every cluster (see the proof in Appendix~\ref{sec:appD}).
\begin{equation} \label{eq:universal} 
\frac{B_\mathrm{c}}{B_\mathrm{0}}=V \frac{2}{3}\left(\frac{r_s}{R}\right)^3.
\end{equation}

This sum should not be understood as implying that all dipoles within a given cluster contribute equally. Instead, their relative contributions are determined by the factor $\left[\cos(2\theta_i) + 3\right]$. Dipoles located on the $x$-axis contribute with a relative strength of 1, whereas those lying in the plane $(x=0)$ contribute with a strength of 0.5. The factor 2/3 is a weighted average between these two extremes.
Note that the  maximal field is independent of the orientation of the cluster, but the curvature of the field in the center is not. 

A particularly noteworthy aspect of Eq.~(\ref{eq:universal}) is its practical applicability. With $B_\mathrm{0}=2/3 B_\mathrm{R}$ it reads
\begin{equation}
\frac{B_\mathrm{c}}{B_\mathrm{R}}=V \frac{4}{9}\left(\frac{r_s}{R}\right)^3.
\label{eq:C3}
\end{equation}
Here, the constant $r_s$ can be replaced by a variable diameter $D_\mathrm{s}$ of a spherical magnet. The resulting expression
\begin{equation}
\frac{B_\mathrm{c}}{B_\mathrm{R}}=\frac{V}{18}\left(\frac{D_\mathrm{s}}{R}\right)^3,
\label{eq:C4}
\end{equation}
then yields the correct value for arbitrary sphere sizes, subject to the obvious geometric constraint imposed by the edge length $a$, i.~e. $D_\mathrm{s}<a/2$.
From an experimental perspective, cubes are often preferred over spheres. A cube with side length $D_\mathrm{c}$ has a dipole moment larger by a factor of $6/\pi$ than that of a sphere with diameter $D_\mathrm{s}$. Consequently, this leads to
\begin{equation}
\frac{B_\mathrm{c}}{B_\mathrm{R}}\approx \frac{V}{3\pi}\left(\frac{D_\mathrm{c}}{R}\right)^3.
\end{equation}
For cubes of side length 20~mm arranged in an icosahedral geometry with radius 41.85~mm, the exact theoretical value deviates by less than $0.2\,\%$ from this approximation. This level of accuracy is comparable to that of the continuum-theory approximation, which differs by less than $0.3\,\%$ from the exact result in this case. 

It is also interesting to compare the exact theory, Eq.~(\ref{eq:C4}), with the estimation based on the continuum theory. Taking the ratio between the term in Eq.~(\ref{eq:C4}) and the term derived with continuum theory of spherical shells Eq.~(\ref{eq:sph_th}), including the volume correction Eq.~(\ref{eq:vol_ratio}) gives:
\begin{equation}
\frac{B_\text{c}^\text{exact}}{B_\text{c}^\text{cont}}=\frac {V \frac{4}{9}\left(\frac{D_s}{2R}\right)^3}{\frac{4}{3}\ln\left(\frac{R_\mathrm{out}}{R_\mathrm{in}}\right) v_r}
\label{eq:ratio}
\end{equation}
which can be expressed with $v_\mathrm{r} = \frac{V \cdot D_\mathrm{s}^3}{D_\mathrm{out}^3-D_\mathrm{in}^3}$ and 
 $\eta=\frac{D_\mathrm{s}}{2R}$ as:
\begin{equation}
\frac{B_\text{c}^\text{exact}}{B_\text{c}^\text{cont}}=\frac {\left(1+\eta\right)^3-\left(1-\eta\right)^3}
{3 \ln\left(\frac{1+\eta}{1-\eta}\right)}
\label{eq:xxx}
\end{equation}
which yields, to the lowest order in $\eta$,
\begin{equation}
\frac{B_\text{c}^\text{exact}}{B_\text{c}^\text{cont}}=1-\eta^4/5 + \cdots
\end{equation}
i.~e.\ both terms coincide for sufficiently small spheres. The highest value of the ratio $\eta$, and thus the largest estimation error, is obtained for a tetrahedron with spheres in contact. This configuration yields $\eta=2/\sqrt{6}$, as follows from the properties of Platonic solids listed in Sec.~II of the Supplemental Material. Even in this worst case, the estimation error is only $13\,\%$, as illustrated in Fig.~\ref{fig:2}.

\section{\label{sec:appD} Proof of Eq.~(\ref{eq:universal})}
It is particularly striking that this term exhibits complete universality across all regular and semiregular clusters. A general proof could not be identified; therefore, each individual cluster was verified using computer algebra. Within this approach, certain shortcuts could, in principle, be employed. For example, once the result has been established for the smallest cluster—the tetrahedron—independently of its orientation, an additional calculation for the cube is unnecessary, since its eight vertices form a compound of two tetrahedra. A similar argument applies to the dodecahedron, which can be decomposed into five tetrahedra. However, such shortcuts do not lead to a substantial reduction in computational complexity. Although a more elegant mathematical proof of Eq.~(\ref{eq:universal}) is likely to exist, none was found.

Because the icosahedron plays a central role in the present work, an outline of the analytical proof for this cluster is provided. The twelve vertices can be grouped into three mutually perpendicular golden rectangles (with the golden side ratio $\varphi$). As a result, only three distinct angles $\theta$ appear in Eq.~(\ref{eq:C1}), namely $\tan^{-1}(\varphi)$,  $\tan^{-1}(1/\varphi)$, and $\pi/2$, each of these angles occurs four times. With 
\begin{equation*}
\begin{array}{lcl}
\cos\!\left(2\tan^{-1}(\varphi)\right)        & = & -\dfrac{1}{\sqrt{5}},\\
\cos\!\left(2\tan^{-1}\!\left(\frac{1}{\varphi}\right)\right) & = & \dfrac{1}{\sqrt{5}},\\
\cos\!\left(2\frac{\pi}{2}\right)             & = & -1,
\end{array}\end{equation*}
the terms in the sum of Eq.~(\ref{eq:C1}) are then
\begin{equation*}
4\left(3-\frac{1}{\sqrt{5}}\right)+4\left(3-\frac{1}{\sqrt{5}}\right)+4\left(3-1\right) = 4\cdot8=4 V \frac{2}{3},
\end{equation*} with $V=12$, which leads to Eq.~(\ref{eq:universal}).

\section{\label{sec:appE} Symmetries and the order of the central saddle point}

To relate the symmetries of the magnetic cluster to the power laws of the ensuing  magnetic field near the center, the magnetic scalar potential $\Phi_\mathrm{m}$ is used because it satisfies the Laplace equation
\begin{equation}
\nabla^2 \Phi_\mathrm{m} = 0,
\end{equation}
and can therefore be expressed as a series expansion in spherical harmonics. In general, the potential takes different forms inside and outside the source region. For the region \emph{outside} the sources ($r>R$):
\begin{equation}
\Phi_m^{(\mathrm{ext})}(r,\theta,\phi) =
\sum_{l=0}^{\infty}\sum_{m=-l}^{l}
\frac{B_{lm}}{r^{l+1}}\,Y_l^m(\theta,\phi),
\label{eq:decay}
\end{equation}
where the terms decay with distance as \(r^{-(l+1)}\).

For the region \emph{inside} the sources ($r<R$):
\begin{equation}
\Phi_m^{(\mathrm{int})}(r,\theta,\phi) =
\sum_{l=0}^{\infty}\sum_{m=-l}^{l}
A_{lm}\,r^{l}\,Y_l^m(\theta,\phi),
\label{eq:increase}
\end{equation}
where the terms remain finite at the origin. Here, \(A_{lm}\) and \(B_{lm}\) are coefficients determined by the boundary conditions on the reference sphere \(r=R\), and \(Y_l^m(\theta,\phi)\) denote the spherical harmonic functions.

\begin{figure}[b]
\includegraphics[width=.48\textwidth]{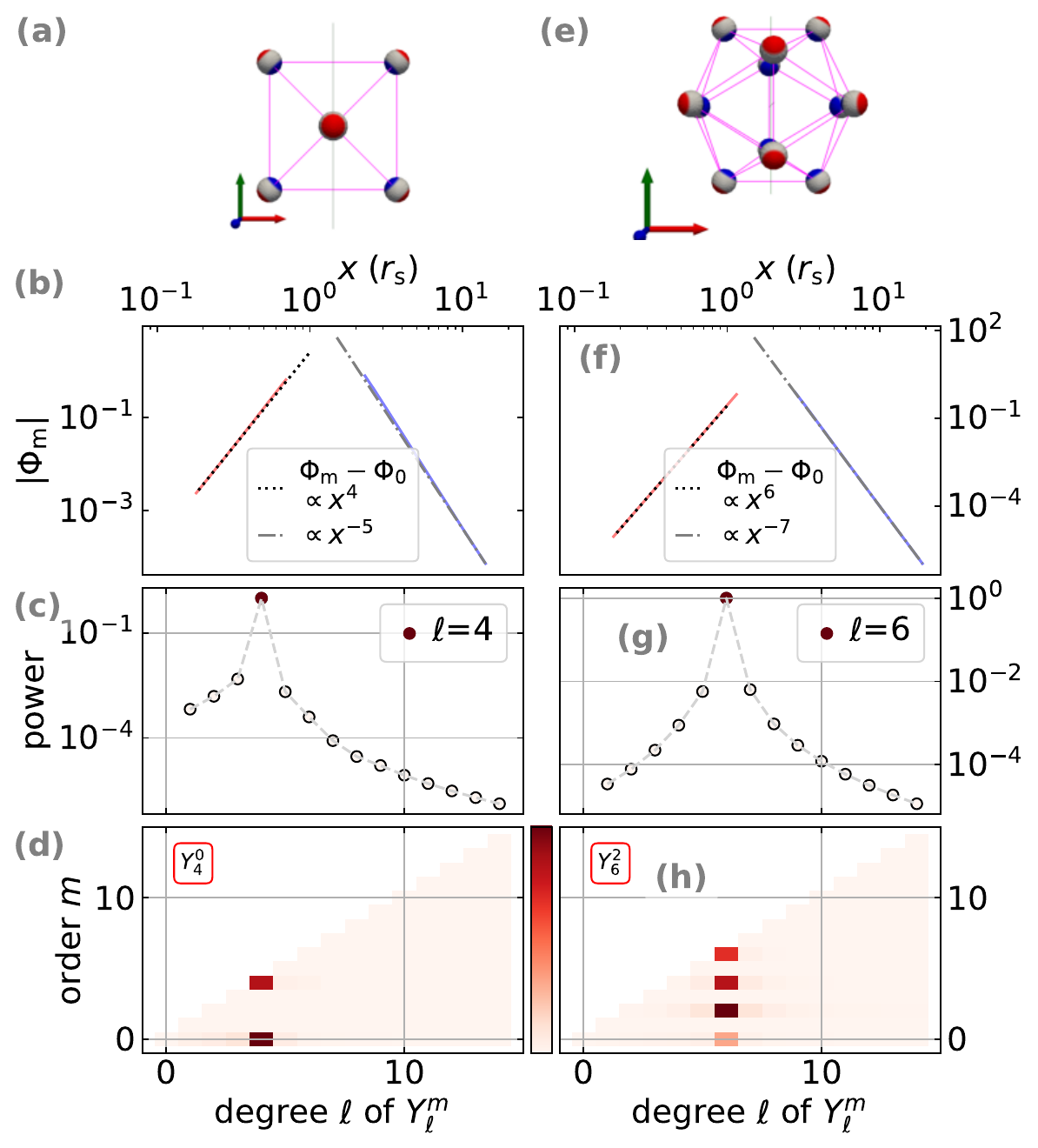}
\caption{Spherical harmonic analysis of the scalar magnetic potential $\Phi_\mathrm{m}$ for an octahedral and an icosahedral cluster in ``star'' 
configuration. (a,e) Configurations, similar to Fig.~\ref{fig:Tetrahedron}(a). (b,f) Asymptotic behavior, similar to  Fig.~\ref{fig:Tetrahedron}(b, c). (c, g) Power spectra from spherical harmonic analysis. All squared $m$-terms belonging to one degree $\ell$ are summed up. (d, h) Color coded spectra, specified according to the order and degree of the modes. The maximum is named in the red legend. See text for more details.}
\label{fig:spec_pot_star}
\end{figure}

This behavior is illustrated by a spherical-harmonic analysis of the scalar magnetic potential $\Phi_\mathrm{m}$ for two representative clusters, shown in Fig.~\ref{fig:spec_pot_star}. For this illustration, the symmetry imposed by the vertex positions is preserved by choosing the dipole moments $\mathbf{m}$ to be parallel to $\mathbf{r}$, corresponding to a star-like orientation. The analysis is performed on the surface of a sphere with radius $0.2\,r_\mathrm{s}$.

For the octahedral cluster shown in Fig.~\ref{fig:spec_pot_star}(a), the lowest nontrivial modes occur at $\ell=4$ \cite{Wojciechowski1997}. According to Eq.~(\ref{eq:decay}), this corresponds to a decay of the potential $\propto\,r^{-5}$. Near the center of the cluster, the potential increases $\propto\,r^{4}$ relative to its central value $\Phi_{0}$, as expressed by Eq.~(\ref{eq:increase}). Both power laws are visible in Fig.~\ref{fig:spec_pot_star}(b), where they appear as the asymptotic behaviors of the numerically calculated red and blue curves.

The icosahedral geometry restricts the lowest allowed spherical-harmonic degree of the potential to a higher order, namely $\ell=6$ \cite{Cohan1958}. With this difference, the discussion of Fig.~\ref{fig:spec_pot_star}(e--h) is entirely analogous to that of the panels (a--d). 

\begin{figure}[hbt]
\includegraphics[width=.48\textwidth]{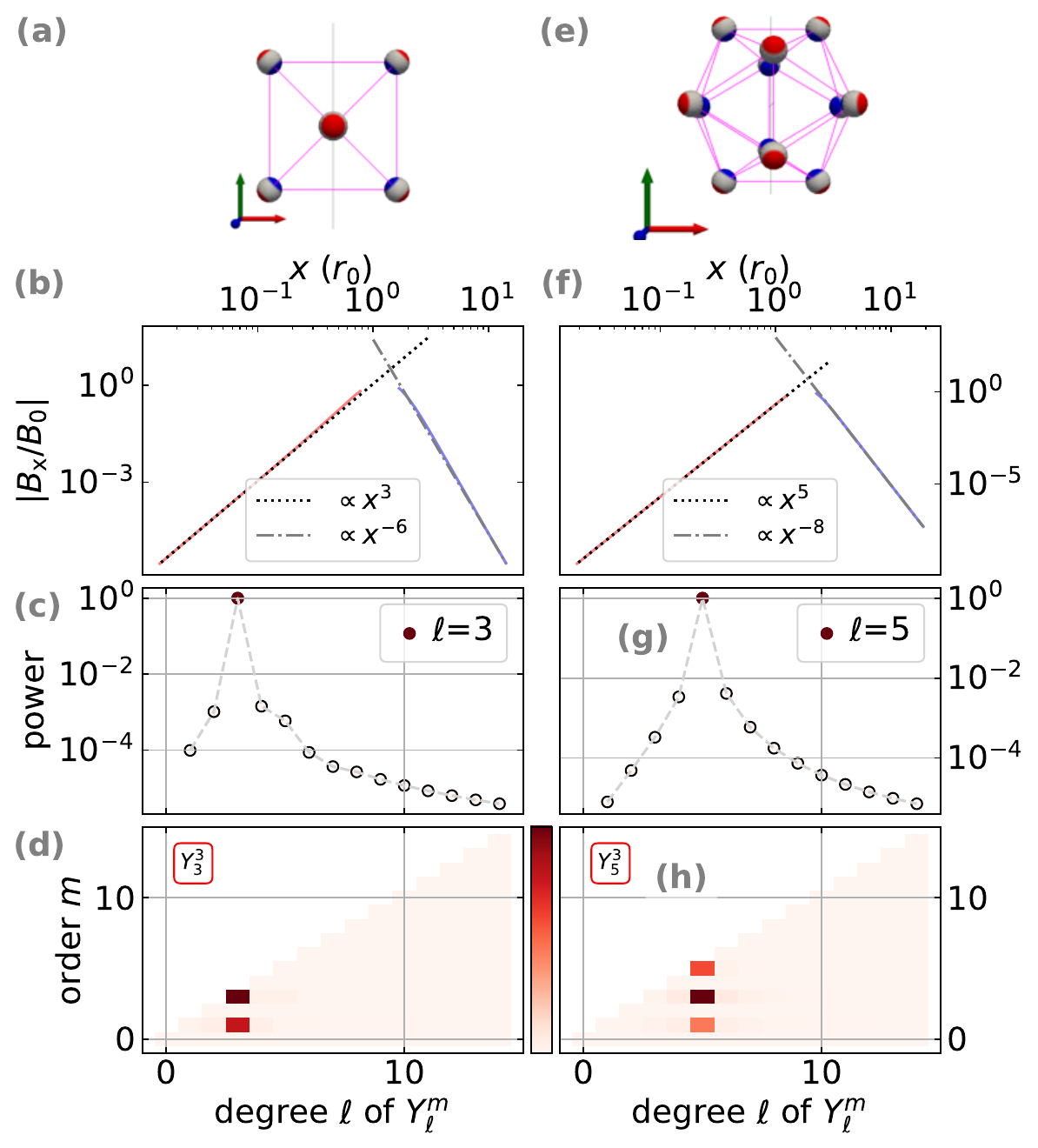}
\caption{Spherical harmonic analysis of $B_x(r=0.2r_\mathrm{s})$ for an octahedral and an icosahedral cluster in "star" configuration. Same presentation as Fig.~\ref{fig:spec_pot_star}.}
\label{fig:spec_star}
\end{figure}

Taking the gradient of the potential to obtain the field is expected to generate spherical harmonics with degrees $\ell \pm 1$. In the panels Fig.~\ref{fig:spec_star}(b) and (g) the mode with the value of $\ell-1$  dominates.

\begin{figure}[ht]
\includegraphics[width=.48\textwidth]{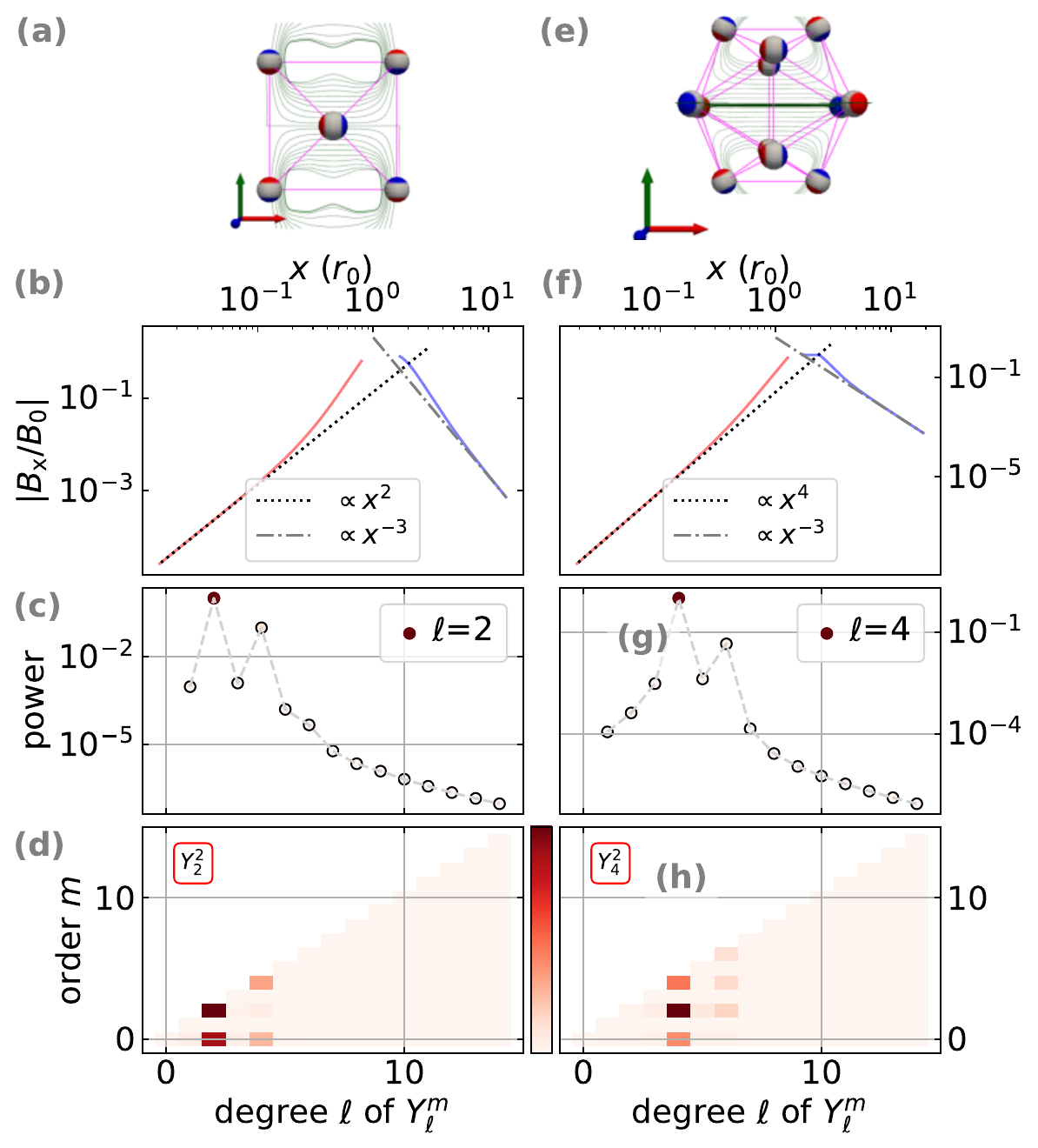}
\caption{Spherical harmonic analysis for an octahedral and an icosahedral cluster in Halbach configuration. Same presentation as Fig.~\ref{fig:spec_pot_star}.}
\label{fig:spec_Halb}
\end{figure}
The Halbach condition in Eq.~(\ref{eq:Halbach}) imposes an azimuthal dependence characterized by $\ell=1$, which leads to additional harmonic components in neighboring bands of the field, as shown in Fig.~\ref{fig:spec_Halb}. In the magnetic field of the octahedral cluster, the modes with $\ell=2$ and $\ell=4$ appear, while for the icosahedral cluster the modes with $\ell=4$ and $\ell=6$ are present in the spherical-harmonic analysis. Near the center of the icosahedron, the magnetic field exhibits a quartic ($r^4$) spatial dependence set by the lowest-degree mode, which provides the leading-order contribution.

As a side note, the focused orientation of the magnets defined by Eq.~(\ref{eq:focused}), which maximizes the magnetic field strength, contains terms with $\ell>1$ (see Fig.~2 of Ref.~\cite{RehbergBlümler2025} for an illustrative comparison of the Halbach and focused configurations). As a consequence, this modulation leads to leakage into additional harmonic bands, and the resulting magnetic field includes a mode with $\ell=2$. This contribution reduces the order of the central saddle point to second order, rendering the focused configuration with its maximal field less favorable in terms of field homogeneity.

In summary, flat fourth-order saddle points arise as a consequence of the suppression of magnetic scalar-potential modes with $\ell<6$, imposed by the icosahedral symmetry.

\bibliography{references}
\end{document}